\documentclass[twocolumn,superscriptaddress,floatfix,preprintnumbers, nofootinbib]{revtex4-2} 
\pdfoutput=1
\usepackage[colorlinks=true,breaklinks=true]{hyperref}
\usepackage[normalem]{ulem}
\usepackage[utf8]{inputenc}
\hypersetup{allcolors=[rgb]{0.0 0.0 0.6},linkcolor=[rgb]{0.75 0.05 0.05}}
\usepackage{amsmath,amssymb}
\usepackage{epsfig}  
\usepackage{graphicx}   
\usepackage{slashed}       
\usepackage{tikz}
\usepackage{color}
\usepackage{url}
\usepackage{color}
\usepackage{multirow}
\usepackage{comment}
\usepackage{amssymb}
\usepackage{bm}
\usepackage{makecell}
\usepackage{enumitem}
\usepackage{diagbox}
\setitemize{itemindent=25pt}

\hypersetup{
colorlinks,
    linkcolor={blue!50!black},
    citecolor={blue!50!black},
    urlcolor={blue!80!black}
}

\newcommand{\beq}{\begin{equation}}
\newcommand{\eeq}{\end{equation}}

\newcommand{\fermi}{\textit{Fermi}}
\newcommand{\sky}{\texttt{skyFACT}}

\definecolor{forestgreen}{rgb}{0.13, 0.545, 0.13}

\allowdisplaybreaks

\bibpunct{[}{]}{,}{n}{}{,}

\newcommand{\crocker}{\hyperlink{cite.Crocker:2022aml}{Crocker2022}}

\begin{document}

\preprint{LAPTH-044/24}

\title{No evidence for gamma-ray emission from \\the Sagittarius dwarf spheroidal galaxy}

\author{Christopher Eckner}
\email{ceckner@ung.si}
\affiliation{Center for Astrophysics and Cosmology, University of Nova Gorica, Vipavska 11c, 5270 Ajdov\v{s}\v{c}ina, Slovenia}
\affiliation{LAPTh, CNRS, F-74000 Annecy, France}
\affiliation{LAPP, CNRS, F-74000 Annecy, France}

\author{Silvia Manconi}
\email{manconi@lapth.cnrs.fr}
\affiliation{LAPTh, CNRS, F-74000 Annecy, France}

\author{Francesca Calore}
\email{calore@lapth.cnrs.fr}
\affiliation{LAPTh, CNRS, F-74000 Annecy, France}

\smallskip

\begin{abstract}
More than a decade ago, the Large Area Telescope aboard the \textit{Fermi} Gamma-ray Space Telescope unveiled the existence of two gigantic gamma-ray lobes known as the \textit{Fermi} bubbles. While their origin is still unknown, various studies identified intricate spectral and morphological structures within the bubbles. One peculiar region, the cocoon, has recently been associated with gamma-ray emissions from the Sagittarius dwarf spheroidal (Sgr) galaxy.
We assess the validity of this claim through adaptive-template fitting and pixel-count statistical methods. Our approach introduces a substantial advancement in data interpretation by enabling a data-driven optimisation of astrophysical background models, thereby reducing the impact of background mis-modelling.
We do not find evidence for gamma-ray emission from the Sgr region at the level obtained in previous work. We demonstrate that there is no pronounced difference between the source population located within the cocoon region and a reference region at similar latitudes. We examine the hypothesis that a millisecond pulsar population in Sgr causes the putative signal via dedicated simulations finding that it is unlikely that we may have detected their presence in Sgr. 
\end{abstract}

\maketitle

\section{Introduction}
\label{sec:introduction}
The \textit{Fermi} Bubbles (FBs) are large, bipolar structures emanating from the Galactic centre, observed in gamma rays extending roughly 10 kpc above and below the plane of the Milky Way \cite{Su:2010qj}. The so-called ``cocoon'' region is a smaller, more localised area within the Southern FB, characterised by enhanced gamma-ray emission. The excess emission has been proposed to represent the cocoons of an old jet activity at the Galactic centre, see the review \cite{2024A&ARv..32....1S} and references therein.

It was noticed in \cite{Crocker:2022aml}---from now on called `\crocker'---that there exists a spatial coincidence between the Sagittarius (Sgr) stream and the FBs, 
and, more specifically, between the core of the Sgr dwarf spheroidal galaxy (Sgr dSph) and the FBs' cocoon.

The Sgr dSph is one of the Milky Way's nearest and most massive satellite galaxies. Located at a distance of approximately 26.5 kpc, it has a mass of about 10$^8$ M$_\odot$ \cite{1994Natur.370..194I, 2003ApJ...599.1082M}, is positioned at Galactic coordinates ($\ell$, $b$) = (5.6$^\circ$, -14$^\circ$) and it is characterised by a prominent stellar stream resulting from its multiple passages through the Milky Way disc, which has stretched and distorted its structure over time \cite{2001ApJ...551..294I}. The galaxy hosts about 11 known globular clusters, which are among the oldest star clusters in the Milky Way \cite{2010MNRAS.404.1203F}. 

By analysing data taken by the Large Area Telescope (LAT) aboard the {\it Fermi} satellite, \crocker~found evidence that the so-called ``cocoon'' region of the FBs originates directly from the Sgr dSph in the background of this extensive hour-glass-shaped structure. They attributed this to the cumulative emission of an old population of millisecond pulsars (MSPs), a well-known class of 
Galactic gamma-ray emitters, within the Sgr dSph.

In the present work, we first aim to examine the evidence for intrinsic emission of the Sgr dSph in light of analysis tools that go beyond the scope of traditionally employed template-based maximum likelihood fits, as the ones applied in \crocker.
The robustness of template fits has been questioned in different contexts such as the GeV gamma-ray excess towards the Galactic centre uncovered by \fermi LAT (see e.g.~\cite{Bartels:2017vsx,Calore:2021jvg,Buschmann:2020adf,Caron:2022akb}).
The main culprit of template-based studies is their limited freedom to account for uncertainties of the spatial morphology of the considered gamma-ray emission components. Hence, results may be inadvertently driven by background mis-modelling. In contrast, we will employ adaptive template fitting as implemented in the analysis software \sky~\cite{Storm:2017arh}. \sky~introduces a framework that joins the functionality of traditional template-based fits as well as image reconstruction algorithms to allow for improved flexibility of the general structure of the compiled gamma-ray emission model (see Sec.~\ref{sec:skyfact} for details).

Secondly, we will put at test the interpretation of the signal in terms of an MSP population by looking for 
point-like sources (PLSs) with gamma-ray fluxes below the detection threshold of LAT catalogues. 
To this end, we will combine \sky~with photon count statistical techniques as implemented within the  
the 1-point probability distribution function (1pPDF) method~\cite{2011ApJ...738..181M, Zechlin:2015wdz}.
By exploiting the photon-count statistics of gamma rays, the collective flux distribution of sources can be measured down to at least one order of magnitude lower fluxes with respect to standard techniques used to build source catalogues~\cite{2011ApJ...738..181M}. 
Methods using this strategy have been developed and efficiently used to measure the properties of faint, unresolved sources in \fermi-LAT data at high Galactic latitudes and in the inner Galaxy, see e.g.~\cite{Lee:2015fea,Mishra-Sharma:2016gis,Zechlin:1,Zechlin:2,Zechlin:3,Lisanti:2016jub,Manconi:2019ynl,Calore:2021jvg}.
To our knowledge, photon-count statistics techniques are used here for the first time to probe the source-count distribution of a putative extended, extragalactic object such as the Sgr dSph. 
We expect the 1pPDF technique to shed light on the emission towards the cocoon region from a different perspective. 
We will inspect the cocoon region's source count distribution, i.e.~the abundance of PLSs as a function of their emitted gamma-ray flux integrated over a specific energy range. The combination of \sky~and 1pPDF techniques allows us to model bright and dim PLSs, while reducing background mis-modelling, as demonstrated in~\cite{Calore:2021jvg}. 

Finally, we will scrutinise, through dedicated Monte Carlo simulations, if and under which conditions the putative Sgr's MSP population 
claimed by \crocker~can be detected in our analysis framework.

We provide below a summary of our main results:
\begin{itemize}
\item \textbf{No statistically significant evidence for gamma-ray emission from Sgr dSph.} When allowing for more flexibility in the fit, by including spatial and spectral re-modulation of diffuse backgrounds with \sky, the statistical evidence for gamma-ray emission of the Sgr dSph is strongly reduced, if not completely washed out. Accordingly, the residuals are reduced to less than 10\% and do not present structures. This remains true even when testing several modelling 
and analysis systematics.
\item \textbf{No evidence for additional point source populations in the Sgr dSph sky region.} The 1pPDF results do not reveal any significant difference in the averaged source-count distribution in the Sgr dSph region with respect to a control region where no emission from the dwarf is expected.
We therefore do not find evidence for the need for an additional source
population in this region with respect to the disc and extragalactic ones. 

\item \textbf{Simulations and injection tests support robustness of our \sky-1pPDF approach.} We verify with simulations and injection tests that, if there had been gamma-ray emissions following a diffuse 
Sgr dSph morphology at the magnitude that \crocker~reported, we would have seen an unmistakably strong signal in the LAT dataset analysed, even in the \sky–approach.
\end{itemize}

The paper is organised as follows: 
In Sec.~\ref{sec:analysis}, we present data selection, and fitting methodologies, 
namely the \sky~(Sec.~\ref{sec:skyfact}) and 1pPDF (Sec.~\ref{sec:1p-pdf}) formalisms.
Sec.~\ref{sec:sgr_physics} outlines how the emission in the analysed region of interest (ROI) is broken down into several model components, for both the \sky~and 1pPDF analyses.
Results of the \sky~analysis are reported in Sec.~\ref{sec:results-skyfact}, 
while we dedicate Sec.~\ref{sec:results-1ppdf} to the 1pPDF main findings. 
Finally, in Sec.~\ref{sec:simulations}, we extensively study if and how 
we could have missed a signal compatible with \crocker's results. 
We finally draw our conclusions in Sec.~\ref{sec:conclusion}.
Our work is complemented by a series of Appendices~\ref{app:astro-model-details}, \ref{app:baseline-runs}, and \ref{app:modelA-runs}.

\section{Analysis framework}
\label{sec:analysis}
Our study employs two different approaches targeted at complementary scientific objectives, and 
articulated into the two analysis frameworks described below. A summary of the prepared datasets' specifications is given in Tab.~\ref{tab:data_selections}.

\subsection{\fermi-LAT data selection}
\label{sec:data-selection}
We work with roughly 12 years of LAT Pass8 data collected from the 4th of August 2008 to the 2nd of August 2020.
We require zenith angles of less than $90^{\circ}$ to reduce the contamination by photons from the Earth's limb. The additional event quality cuts (\texttt{DATA\_QUAL>0 \&\& LAT\_CONFIG==1}) are likewise applied. All selection, cleaning, manipulation and simulation of \fermi-LAT~data is conducted via the \textit{Fermi} Science Tools\footnote{\url{https://github.com/fermi-lat/Fermitools-conda}} (version 2.2.0) \cite{2019ascl.soft05011F}.

\begin{figure}
\centering\includegraphics[width=\linewidth]{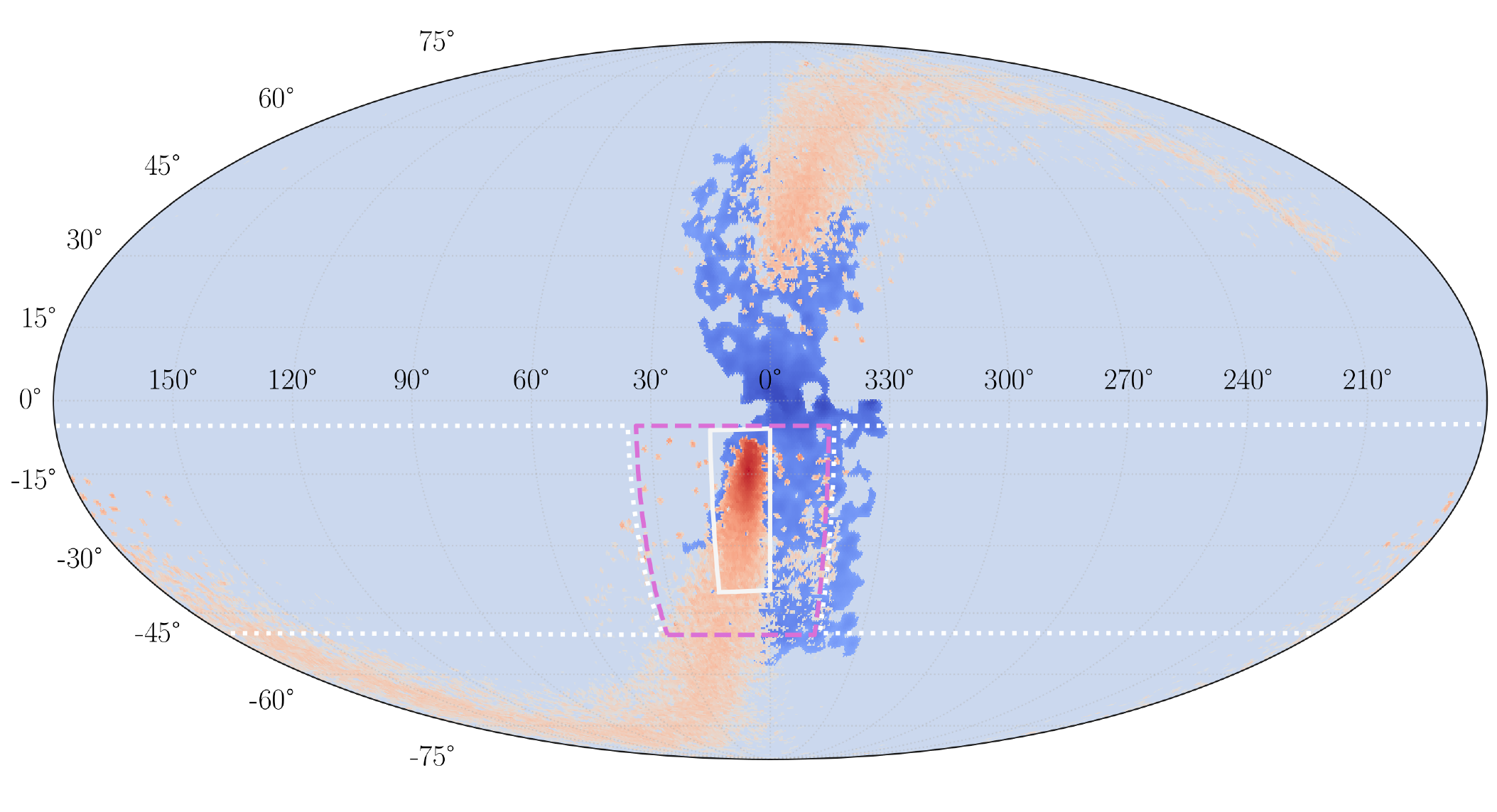}
    \caption{ROIs analysed in this work, as represented in Mollweide projection.  The pink dashed line underlines the region used for the \sky~analysis of the Sgr dSph (cf.~Fig.~\ref{fig:sgr-templates} in Cartesian projection). The smaller, white region is the one used for the 1pPDF analysis.  The region delimited by white dotted lines is  ``Anti-Sgr dSph'' region used to compare the source-count distribution  at the same latitudes of the Sgr dSph. The FBs template (taken from \cite{Fermi-LAT:2017opo}, in blue) and the Sgr dSph density map from the early \textit{Gaia} data release 3 \cite{Ramos2022:abc} (red) are overlaid for comparison. \label{fig:rois} }
\end{figure}

Through the \sky~analysis we first aim at scrutinising the results reported in \crocker~so that we adopt a \fermi-LAT data selection very close to its specifications but updated to a longer time interval. 
We consider events within a square ROI of $50^{\circ}\times50^{\circ}$ centred on the sky position $(\ell, b) = (10^{\circ}, -25^{\circ})$ in Galactic longitude and latitude, namely the centre of the ROI in \crocker.
This ROI is shown in Fig.~\ref{fig:rois} delimited by a pink dashed line.
While we use the full ROI specified in Tab.~\ref{tab:data_selections} in this analysis to define the initial model templates and to select the set of localised gamma-ray sources, we reduce the fitted region around the centre of the projected binned dataset to a $40^{\circ}\times40^{\circ}$ square. This way, we still incorporate gamma-ray emission from the larger ROI caused by potential spill-over effects due to the point-spread function (PSF) convolution. We also avoid the need to precisely model and fit the gamma-ray emission along the Galactic plane at Galactic latitudes $b > -5^{\circ}$. We note that this ROI definition was also used in \crocker.
Photons satisfying the reconstruction criteria of the event class \texttt{P8R3\_ULTRACLEANVETO\_V3} for \texttt{FRONT+BACK} type events are part of the dataset. We consider gamma rays with energies from 500 MeV to 177.4 GeV.  Following \crocker, we bin the prepared dataset into 17 logarithmically spaced energy bins but refrain from resuming the last four bins into two macro energy bins. As concerns the spatial binning, we use square pixels of $0.25^{\circ}\times0.25^{\circ}$, while the native binning in \crocker~is $0.2^{\circ}\times0.2^{\circ}$. However, the resolution of the employed initial set of templates does not impact the final results in the \sky~framework as we explain in Sec.~\ref{sec:skyfact}. The increased exposure is purposefully chosen to match the period adopted in the \fermi-LAT collaboration's 3rd release (DR3) of the 4FGL gamma-ray source catalogue \cite{Fermi-LAT:2019yla, Fermi-LAT:2022byn}.

To increase the photon statistics available to the 1pPDF method, we generate a second dataset that is based on the event class \texttt{P8R3\_SOURCEVETO\_V3}. In fact, up to 10 GeV, this event class employs the same selection criteria as the \texttt{SOURCE} event class, which is the recommended class for most analyses dealing with point sources or slightly extended sources. Above 10 GeV, the \texttt{SOURCEVETO} event class is almost identical to the \texttt{ULTRACLEANVETO} class. We restrict the dataset to energies between 2 to 5 GeV and to an ROI embedding the expected peak of the Sgr dSph stellar density, as illustrated in Fig.~\ref{fig:rois}.  An additional ROI covering the Galactic plane, excluding the longitudes around the Sgr dSph (white dotted ROI in Fig.~\ref{fig:rois}) is used for comparison. We adopt a HEALPix pixelisation \cite{2005ApJ...622..759G} with $N_{\mathrm{side}} = 7$ corresponding to a pixel size of about $0.5^{\circ}$. Finally, we restrict the events to the \texttt{PSF3} event type following \cite{Zechlin:2015wdz,Calore:2021jvg}.

\begin{table*}[t]
\begin{centering}
{\renewcommand{\arraystretch}{1.5}
\begin{tabular}{l|p{4cm}|p{4cm}}
\Xhline{5\arrayrulewidth}
{Data Set} & \sky~analysis & 1pPDF analysis \\
\hline
{Reconstruction algorithm} & \multicolumn{2}{c}{{Pass 8}}\\
{Time interval} & \multicolumn{2}{c}{{\makecell{12 years (4th August 2008  - 2nd August 2020)}}}\\
{Event class} & \texttt{ULTRACLEANVETO} & \texttt{SOURCEVETO}\\
{Event type} & \texttt{FRONT+BACK} & \texttt{PSF3}\\
{Energy range}  & {500 MeV - 177.4 GeV} & {2 GeV - 5 GeV}\\
{Projection} & plate carrée (CAR) & HEALPix\\
{Spatial binning} & {$0.25^{\circ}\times0.25^{\circ}$} & {$N_{\mathrm{side}} = 7$ ($\sim0.5^{\circ}$)}\\
{Energy binning} & {17; log-spaced} & {1}\\
{Zenith angle cut} & \multicolumn{2}{c}{{$\leq90^{\circ}$}}\\
{Time cuts filter} & \multicolumn{2}{c}{{\makecell{DATA\_QUAL$>$0 \&\& LAT\_CONFIG==1}}}\\
\Xhline{5\arrayrulewidth}
\end{tabular}
}
\par\end{centering}
\caption{Data selection and preparation specifications utilised in combination with \sky~(see Sec.~\ref{sec:skyfact}) and the 1pPDF (see Sec.~\ref{sec:1p-pdf}) analyses.\label{tab:data_selections}}
\end{table*}

\subsection{\sky~analysis}
\label{sec:skyfact}
\textbf{Adaptive-template fitting.} \sky~supplements the methodology and statistical framework of traditional template-based maximum likelihood fits with image-reconstruction techniques. The full framework of this approach is described in \cite{Storm:2017arh}, as well as partially in publications applying \sky~to science cases like the Galactic centre excess \cite{Bartels:2017vsx, Calore:2021jvg, Manconi:2024tgh, Song:2024iup} or the Andromeda galaxy~\cite{Armand:2021rlr}.

The image-reconstruction part is realised as spatial and spectral re-modulation of all components of the employed gamma-ray emission model. Such flexibility is achieved by introducing a large number of nuisance parameters in the form of bin-by-bin variations that affect either the spatial pixels $p$ or energy bins $b$. The predicted gamma-ray flux $\phi_{pb}$ per bin is consequently given by a tri-linear sum over the components of the considered gamma-ray emission model
\begin{equation}
\label{eq:skyfact-base-eq}
    \phi_{pb} = \sum_{k} \tau_{p}^{(k)}T^{(k)}_{p}\cdot\sigma_{b}^{(k)}S^{(k)}_{b} \cdot \nu^{(k)}\mathrm{.}
\end{equation}
Here, $T^{(k)}_{p}$ refers to the morphology of emission component $k$ in pixel $p$ and $\tau_{p}^{(k)}$ denotes its associated modulation parameter. The analogous construction is assumed for the spectral part of the emission component $S^{(k)}_{b}$ in energy bin $b$, which is varied by the spectral modulation parameter $\sigma_{b}^{(k)}$. Each emission component may be re-normalised by the overall parameter $\nu^{(k)}$. The nuisance parameters $\bm{\tau}$, $\bm{\sigma}$ and $\bm{\nu}$ are taken to be strictly positive (thus, ensuring a physical interpretation) and meant to be varied around unity as a way to account for uncertainties of the morphology and spectrum of the initial gamma-ray emission model. An exception to this general strategy is the treatment of PLSs. Their positions are fixed, i.e.~$T^{(k)}_{p}$ is not part of the model, and only their spectra are re-modulated.

To avoid overfitting or unphysical re-modulation of the components, one wishes to control the \emph{magnitude} of the re-modulation parameters and the \emph{tension} between them. This is achieved by introducing a penalising likelihood function in addition to the standard Poisson likelihood term employed in template-based fits (e.g., \crocker~or \cite{Calore:2014xka, Calore:2014nla, Macias:2019omb, CTA:2020qlo, Calore:2021hhn, Pohl:2022nnd, Eckner:2022swf, Song:2024iup}). While the full details of the penalising part of the likelihood function are given in \cite{Storm:2017arh}, we will limit ourselves here to a summary. The magnitude of the re-modulation parameters $\left(\bm{\tau}, \bm{\sigma},\bm{\nu}\right)$ is restricted via the maximum entropy method \cite{1979MNRAS.187..145S, 1992daia.conf..251S, 1994IAUS..158...91B} and the set of hyper-parameters $\left(\lambda, \lambda^{\prime},\lambda^{\prime\prime}\right)$, respectively. Note that each model component is assigned an associated magnitude hyper-parameter. The tension between spatial and spectral bins is essentially a means to introduce a smoothing scale/correlation length among adjacent bins. The idea is to restrict the freedom of each bin to move independently of its neighbouring bins. In the case of spatial bins, a tension parameter $\eta$ quantifies the tension between the (up to four) nearest-neighbour bins while $\eta^{\prime}$ fulfils the same task for adjacent energy bins. As an example, the spatial and spectral tension scales $\eta$ and $\eta^{\prime}$ are given by $\eta = 1/\delta x^2$ ($\eta^{\prime} = 1/\delta E^2$), where $\delta x$ ($\delta E$) is the allowed variation between neighbouring spatial pixels (energy bins). Setting the tension to zero is thus equivalent to saying that all spatial pixels (energy bins) may vary entirely independently of each other. 

After preparing the flux model $\phi_{pb}$ per bin, we obtain the expected photon counts by convolving the model with the LAT's PSF and multiplying the result by the binned exposure map corresponding to the data selection. Both, PSF and exposure map are derived via the \textit{Fermi} Science Tools. The subsequent fit to \fermi-LAT data relies on the L-BFGS-B algorithm \cite{53712fe04a3448cfb8598b14afab59b3, 10.1145/279232.279236, 10.1145/2049662.2049669} implemented in the \texttt{python3} module \texttt{scipy.optimize} that minimises the sum of Poisson and penalising likelihood function. 

\noindent\textbf{Dissecting the Sgr dSph region with \sky.} We consider a fiducial gamma-ray emission model for the Sgr ROI, comprised of components outlined in Sec.~\ref{sec:sgr_physics}.
We are interested in assessing the impact of \emph{background mis-modelling} on the significance of the Sgr dSph template. 
To this end, we conduct a sequence of fits with the same gamma-ray emission model, i.e.~our {\it baseline} model, with the following rationale: We begin with a template fit, i.e.~$\bm{\tau}^{(k)} \equiv 1$ (as utilised in \crocker) is fixed for all components, and successively allow for spatial re-modulation component by component. Thus, we can infer which components are driving the significance for the Sgr dSph. By default, we adopt the \sky~hyper-parameters of \textsc{run5} of \cite{Storm:2017arh}, which have been identified as a reasonable choice (for fits of the gamma-ray emission along the Galactic plane). We comment on changes to these settings whenever necessary. 

To quantify the significance of the Sgr dSph emission in each step of our systematic scan, we keep its adopted spatial morphology fixed while allowing for full spectral freedom (without any correlation among the energy bins). This is the traditional, template-fit approach. Consequently, we can compute the significance of the component from the minimal value of the total (log-)likelihood function in comparison to a fit whose gamma-ray emission model does not include the Sgr dSph. We create a nested model and employ the log-likelihood ratio test statistic $\mathrm{TS} = 2(\ln{\mathcal{L}_{\mathrm{base+Sgr}}} - \ln{\mathcal{L}_{\mathrm{base}}})$ as the foundation of our statistical interpretation where the baseline model does not incorporate the Sgr dSph contribution. Adding its emission to the baseline model adds 18 degrees of freedom (17 energy bins + one overall normalisation). In such a case, the test statistic is distributed according to a mixture distribution following \citep{Macias:2016nev}
\begin{equation}
    p(\mathrm{TS}) = 2^{-N_{\mathrm{dof}}}\left[\delta(\mathrm{TS}) + \sum_{k = 1}^{N_{\mathrm{dof}}} \binom{N_{\mathrm{dof}}}{k}\chi^2_{k}(\mathrm{TS})\right]\rm{,}
\end{equation}
where $\delta$ is the Dirac distribution, $N_{\mathrm{dof}}$ denotes the number of added degrees of freedom, $\binom{n}{k}$ is the binomial coefficient and $\chi^2_{k}$ refers to a $\chi^2$-distribution with $k$ degrees of freedom. It follows that the significance $\mathcal{Z}$ of the added component under the observation of the test statistic value $\hat{\mathrm{TS}}$ is given by
\begin{equation}
\label{eq:significance-mixture-model}
    \mathcal{Z}(\hat{\mathrm{TS}}) = \sqrt{\mathrm{CDF}^{-1}\!\left(\chi^2_1, \mathrm{CDF}\!\left(p(\mathrm{TS}), \hat{\mathrm{TS}}\right)\right)}\rm{,}
\end{equation}
where $\mathrm{CDF}(f, x)$ denotes the cumulative distribution function of $f$ at $x$ and CDF$^{-1}$ refers to its inverse.

\subsection{1pPDF analysis}
\label{sec:1p-pdf}
Photon count statistical methods permit to measure the collective source-count distribution, $dN/dS$, of sources emitting a gamma-ray flux $S$ as integrated in a given energy bin and region in the sky. This is done by a statistical analysis of the probability distribution $p_k^{(p)}$ of the photon counts $k^{(p)}$ in each pixel $p$ of a pixelised map. 
Indeed, different classes of gamma-ray sources are expected to contribute to $p_k^{(p)}$ with photons following different statistics. Truly diffuse, isotropic emissions are expected to contribute to $p_k^{(p)}$ with photons following a Poissonian distribution. Point sources and structured Galactic diffuse emissions instead provide non-Poissonian contributions to the $p_k^{(p)}$. By investigating the $p_k^{(p)}$ of photon counts, one can thus extract statistical information of the different source classes contributing to observed gamma rays. 
The core formulation of the method can be found in Refs.~\cite{2011ApJ...738..181M,Zechlin:1}, which we refer to for any detail, such as the treatment of the PSF and exposure of \fermi-LAT data. 
We note that recently a number of machine-learning-based methods have been developed to attempt the same goal, letting the algorithm learn the underlying pixel-by-pixel correlations \cite{Caron:2017udl,List:2020mzd,2021PhRvD.104l3022L,Mishra-Sharma:2021oxe,Amerio:2023uet}. These however could face a reality gap when trained on non-realistic simulations of the gamma-ray sky \cite{Caron:2022akb}. 

The implementation of the 1pPDF is here used, as introduced in Refs.~\cite{Zechlin:1,Zechlin:2,Zechlin:3}.
Specifically, we refer to Refs.~\cite{Calore:2021jvg,Manconi:2024tgh} in which the 1pPDF has been combined with \sky-optimised templates to minimise diffuse mis-modelling. This hybrid approach hampers the robustness of photon count statistical analyses of \fermi-LAT data when analysing regions of high Galactic diffuse emission, such as the inner Galaxy \cite{Buschmann:2020adf}. 

The $dN/dS$ of the point sources in the ROI, assumed to be isotropic, is modelled by means of a multiple broken power law with two free breaks, leaving free the normalisation, the indices of the broken power law above and below the breaks, and the break positions, for a total of six free parameters.  
In addition to point sources, an isotropic diffuse emission with free normalisation,  and diffuse emission templates as obtained with \sky, including e.g.~Galactic diffuse emission or the putative diffuse emission from the Sgr dSph are  
considered, each coming with an additional free normalisation. 
As done in previous applications, we do not mask bright point sources found in \fermi-LAT catalogues. Our measurement of the $dN/dS$ is thus expected to reproduce the number counts of sources in catalogues while extending the measure down to about one order of magnitude lower fluxes. 
The 1pPDF fit is performed taking into account the spatial morphology of the diffuse templates together with the point sources through a pixel-dependent likelihood function, the L2 method as defined in \cite{Zechlin:1}. 
We sample the posterior distribution using \texttt{Multinest} \cite{2009MNRAS.398.1601F}, and obtain both Bayesian and frequentist results as described in \cite{Zechlin:1}. Specifically, we obtain prior-independent, frequentist maximum likelihood parameter estimates by means of one-dimensional profile likelihood by using the final posterior samples. Instead, model comparison is performed using the nested sampling global log-evidence $\ln(\mathcal{Z})$. 

\section{Modelling the gamma-ray sky towards the Sgr dSph}
\label{sec:sgr_physics}

\subsection{\sky~model and templates}
\begin{figure}
\centering\includegraphics[width=1\columnwidth]{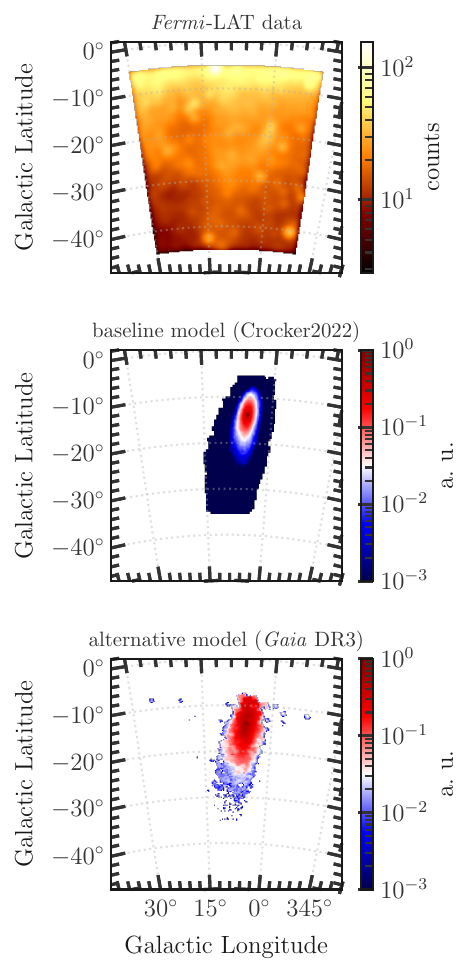}
    \caption{
    (\emph{Upper}:) \textit{Fermi}-LAT data in the Sgr ROI from 1 to 4 GeV smoothed with an 0.8$^{\circ}$ Gaussian kernel to reflect the PSF size at about 1 GeV. (\emph{Middle}:) Baseline template adopted from \protect\crocker~following the catalogue of Sgr member candidates \protect\cite{2020MNRAS.497.4162V} identified in the \emph{Gaia} satellite's Data Release 2. (\emph{Lower}:) Alternative Sgr dSphs density map derived from the catalogue of Sgr stream candidate member stars found in the early Data Release 3 of \emph{Gaia} \protect\cite{Ramos2022:abc}. The density maps are normalised to unity resulting in arbitrary units. The colour scale displays logarithmic values. 
 \label{fig:sgr-templates}} 
\end{figure}

The selected ROI around the FBs' cocoon region is moderately extended requiring a gamma-ray emission model with adequate freedom to capture the different physical processes relevant in the GeV energy range as exemplified in the upper panel of Fig.~\ref{fig:sgr-templates} displaying the LAT data from 1 to 4 GeV in the ROI.
Furthermore, it is in our interest not to change the original assumptions and modelling in \crocker~too much. In this way, we guarantee the relative comparability of both studies. Consequently, our gamma-ray emission model will not go beyond those components assumed by the authors of \crocker, namely:
\begin{enumerate}
    \item point-like sources (from 4FGL-DR3),
    \item extended sources (from 4FGL-DR3),
    \item the diffuse isotropic gamma-ray background (IGRB),
    \item emission related to the Sun and the Moon,
    \item diffuse Milky Way foreground (gas, for short),
    \item inverse Compton (IC) emission,
    \item the FBs and
    \item the Sgr dSph,    
\end{enumerate}
for which we provide the adopted spatial and spectral priors with references in Tab.~\ref{tab:model-components}. Note that \crocker~also considered additional components for the Galactic centre excess and Loop I -- an arc-like extended diffuse structure mainly spanning across the northern hemisphere of the gamma-ray sky \cite{Wolleben:2007pq}. We refrain from doing so since we tested explicitly that these contributions to the overall emission are negligible and not preferred by any of the performed fits. 

\begin{table*}[t!]
    \centering
    {\renewcommand{\arraystretch}{1.5}
    \begin{tabular}{p{4cm} p{12cm}}
    \Xhline{5\arrayrulewidth}
         \textsc{Model component} & \textsc{Description} \\\hline
         \textbf{1.~point-like sources}  & spectral prior: power law $E^{-2.5}$; spatial priors (positions): 4FGL-DR3 catalogue version \href{https://fermi.gsfc.nasa.gov/ssc/data/access/lat/12yr_catalog/gll_psc_v30.fit}{\texttt{gll\_psc\_v30.fit}} \cite{Fermi-LAT:2019yla, Fermi-LAT:2022byn} \\
         \hline
         \textbf{2.~extended sources} & spectral prior: 4FGL-DR3 best-fit spectra in \href{https://fermi.gsfc.nasa.gov/ssc/data/access/lat/12yr_catalog/gll_psc_v30.fit}{\texttt{gll\_psc\_v30.fit}} \cite{Fermi-LAT:2019yla, Fermi-LAT:2022byn}; spatial priors (positions / extensions): 4FGL-DR3 as above \cite{Fermi-LAT:2019yla, Fermi-LAT:2022byn} / as \sky~\textsc{run 5}  \cite{Storm:2017arh}\\
        \hline
        \textbf{3.~IGRB} & spectral prior: \textit{Fermi Science Tools} \texttt{iso\_P8R3\_ULTRACLEANVETO\_V3\_v1.txt}; spatial prior: uniform in ROI\\
        \hline
        \textbf{4.~Sun \& Moon} & spectral and spatial prior from templates in \crocker: \href{https://zenodo.org/records/6210967}{10.5281/zenodo.6210967} based on \cite{2012ApJ...758..140A, 2011ApJ...734..116A}\\
         \hline
         \hline
        \textbf{5.~gas} & \\
        $\quad$\emph{(a)} baseline & spectral prior: foreground model A \cite{Fermi-LAT:2014ryh}; spatial prior:(\textsc{h~i}) -- taken from the 2016 release of the HI4PI collaboration \cite{refId0}, H$_2$ --  third annulus (8.0 - 10.0 kpc) of the H$_2$ templates adopted in \crocker \\
        $\quad$\emph{(b)} ModelA & spectral and spatial prior: foreground model A \cite{Fermi-LAT:2014ryh}\\
        $\quad$\emph{(c)} Model0 & spectral prior: foreground model A \cite{Fermi-LAT:2014ryh}; spatial prior: all gas-related plus dust correction templates in \crocker \\
        \hline
        \textbf{6.~IC emission} & \\
        $\quad$\emph{(a)} baseline & spectral and spatial prior: foreground model A \cite{Fermi-LAT:2014ryh}\\
        $\quad$\emph{(b)} Model0 & spectral and spatial prior: all IC templates used in \crocker\\
        \hline
        \textbf{7.~FBs} & \\
        $\quad$\emph{(a)} baseline (flat) & spectral prior: \cite{Fermi-LAT:2014sfa}, spatial prior: uniform following shape obtained in \cite{Fermi-LAT:2014sfa}\\
        $\quad$\emph{(b)} baseline (structured) & spectral and spatial prior: \cite{Fermi-LAT:2014sfa}\\
        $\quad$\emph{(c)} FB2017 & spectral prior: \cite{Fermi-LAT:2017opo}; spatial prior: inpainted version \cite{Macias:2019omb} of template of \cite{Fermi-LAT:2017opo} \\
        $\quad$\emph{(d)} cocoon region & taken from structured baseline FBs template applying the baseline Sgr template (see Fig.~\ref{fig:sgr-templates}) as a binary mask\\
        \hline
        \textbf{8.~Sgr} & \\
        $\quad$\emph{(a)} baseline & spectral prior: power law $E^{-2}$; spatial prior: fiducial Sgr dSph template of \crocker~from stars selected according to \cite{2020MNRAS.497.4162V}\\
        $\quad$\emph{(b)} SgrGaia3 & spectral prior: power law $E^{-2}$; spatial prior: stellar density of associated Sgr stars from the early \textit{Gaia} DR3 (see text) \cite{Ramos2022:abc}\\
        \Xhline{5\arrayrulewidth}
        
    \end{tabular}}
    \caption{Summary of the components selected to define our baseline model setup. The spectral prior per component refers to the quantity $S^{(k)}$ and the spatial morphology to $T^{(k)}$ in the \sky~model definition stated in Eq.~\ref{eq:skyfact-base-eq}. Model components equipped with the tag ``baseline'' are part of our baseline model; other components are later used to explore the robustness of the method to variations in the model composition. 
    }
    \label{tab:model-components}
\end{table*}

We proceed as follows: Each model component listed in Tab.~\ref{tab:model-components} with the tag ``baseline'' becomes part of our \emph{baseline model} iteration. This baseline model is the benchmark for all statistical statements of the \sky-related study. In addition, we will check the robustness of our findings by changing the composition of the benchmark case. To do this, we exchange individual components and replace them with similar flux models whose derivation is based on slightly different assumptions. 
%
Motivated changes in the baseline model's composition concern the brightest components among the considered gamma-ray emitters. In particular, we modify our assumptions about the gas, IC, FBs and Sgr dSph templates.

The crucial ingredient in our gamma-ray emission model is the spatial morphology assumed for the Sgr dSph. In what follows, we describe the baseline and alternative density profiles we employ in this work. A full description of the remaining model components in terms of the physics and motivation is given in Appendix \ref{app:astro-model-details}, where we also show the spatial morphology of the adopted baseline components in Fig.~\ref{fig:baseline-model} after re-modulating them in our final \sky~hyper-parameter setup.

\emph{Modelling the Sgr dSph.} We adopt as baseline the Sgr dSph spatial template employed in \crocker, as shown in the middle panel of Fig.~\ref{fig:sgr-templates}. This template represents the stellar density of the Sgr dSph's core according to a selection of $2.26\times10^5$ stars classified as Sgr candidate members \cite{2020MNRAS.497.4162V} found after searching the \emph{Gaia} satellite's Data Release 2 \cite{2018A&A...616A...1G}. More than 50\% of the stars selected in \cite{2020MNRAS.497.4162V} are red clump stars and can reliably be separated from field stars. The applied selection criteria are such that the obtained set of stars reproduces the observed properties of the core of the Sgr dSph. Sgr's stellar stream is not part of the template. For a discussion of the full Sgr stream in the selected ROI see Sec.~\ref{sec:stream}.

As an alternative model (SgrGaia3), we consider the sample of stars selected and presented in \cite{Ramos2022:abc} using \emph{Gaia}'s (early) third data release. The latter study aims to construct the full Sgr stream from \emph{Gaia} data and not only its core region. However, to derive a template of the Sgr dSph core, we bin the full catalogue of more than $7\times10^5$ candidate member stars in a HEALPix grid with $N_{\mathrm{side}} = 256$ \cite{Gorski:2004by}. Then, we apply an empirical threshold of 18 stars to remove parts of the template that are likely not associated with the dSph's core region. We use the resulting two-dimensional map as an alternative model for the Sgr dSph. The obtained stellar density of Sgr is shown in the lower panel of Fig.~\ref{fig:sgr-templates}. The apparent differences in the density profiles of the two models are due to the different selection of stars in the \textit{Gaia} catalogue. The middle plot exhibits a much denser region towards the core of the Sgr dSph as was intended during the data selection. The bottom plot shows a Sgr profile that is much more extended and shallower towards its core as it incorporates all stars associated with Sgr and its stellar stream. Also note the small islands of non-zero density above the threshold on the number of stars. We checked \textit{a posteriori} by removing these features from the template that they do not drive the significance for Sgr.
Finally, we also tested what happens when a realistic template for Sgr is 
replaced by a simple, geometrical, template for the cocoon emission. 
We generate a cocoon template from the structured FBs template using the baseline Sgr dSph template as a binary mask. This approach cuts out the desired target region within the FBs due to the alignment of Sgr with the cocoon. Hence, the ``Cocoon'' model iteration contains all templates of the baseline setup plus the hence-created cocoon template instead of the Sgr dSph template. For this test, we assume a flat spectral prior for the cocoon template, i.e.~a power law $\sim E^{-2}$ as we did for the Sgr template in our systematic runs.

\subsection{1pPDF model}
In short, the 1pPDF fit includes: a point source distribution modelled with a $dN/dS$,  parametrised as multiple broken power law with two free breaks; the isotropic gamma ray background; a gamma-ray diffuse emission model, which is obtained summing the gas, IC, IGRB, Sun and Moon, and FBs templates as optimised within \sky, that is integrated within 2--5~GeV and enters the 1pPDF fit with an overall free normalisation; the Sgr dSph diffuse template, again integrated  within 2--5~GeV and, when included, entering with a free normalisation between [0.7,1.3]. 

We will perform several fits with the 1pPDF, stressing that each of them will be run with a \sky-optimised 
foreground model. This two-step procedure guarantees full self-consistency between model components in the 1pPDF fit and data-driven foreground optimisation.

\section{Uncovering the gamma-ray emission from the Sgr dSph with \sky}
\label{sec:results-skyfact}

In this section, we report on the results of the \sky-related part of our study. The main scientific objective is to assess the significance of gamma-ray emission from the Sgr dSph in light of known astrophysical background components and their associated uncertainties. We examine the robustness of our analysis by varying the employed hyper-parameters of \sky~per model component, as well as exchanging our baseline model composition with alternative characterisations, which we detailed in Sec.~\ref{sec:sgr_physics}.

\subsection{Direct comparison with the baseline result of \protect\crocker}
\label{sec:skyfact-sgr-crocker-comparison}

As a first step, we take the entire template and component setup provided by the authors of \crocker~and perform a standard template-based fit to our selected \fermi-LAT dataset. In fact, such a fit can be realised in the framework of \sky~by fixing the spatial re-modulation parameters to their initial value of 1 ($\lambda = \infty$) and disabling spatial and spectral smoothing ($\eta = \eta^{\prime} = 0$). We set the penalising likelihood function's contribution to zero by letting $\lambda^{\prime} = \lambda^{\prime\prime} = 0$. Hence, spectrum and overall normalisation are free to vary. Yet, we treat extended and point-like 4FGL-DR3 sources already with the full power of \sky: As concerns PLSs, we allow for up to $20\%$ relative variations of their spectrum (without smoothing) and modest freedom of an overall re-normalisation. Extended sources are initialised as uniform brightness discs and we leave the spatial re-modulation parameters fully unconstrained except for a certain degree of smoothing whereas the spectra are only marginally constrained.

Running this \sky~initialisation yields a direct comparison with the baseline result of \crocker~(modulo the mentioned minor details that distinguish a pure template fit from the one we conduct here with \sky). To stress it, the baseline model of \crocker~contains the data-driven structured FBs template of \cite{Fermi-LAT:2014sfa} as well as the gas and IC templates labelled with ``Model0'' in Tab.~\ref{tab:model-components}. We run this fit twice; with and without the baseline Sgr template to assess the significance of this component over the background model. Numerically, we find likelihood values of $-2\ln{\mathcal{L}}_{\mathrm{base}} = 306916$ and $-2\ln{\mathcal{L}}_{\mathrm{base+SgrStream}} = 306805$ corresponding to a significance of the Sgr dSph of $\mathcal{Z}_{\mathrm{Sgr}} = 8.7\sigma$ over the background following Eq.~\ref{eq:significance-mixture-model}. This result is very close to the reported $8.1\sigma$ in \crocker. Therefore, we confirm the striking evidence of Sgr given this baseline model and the template-fit approach even after updating the dataset to 12 years.

In the following subsection, we address the robustness of this result in the framework of \sky~by allowing for modulations of the spatial and spectral component priors to reduce potential background mis-modelling.

\subsection{Significance of the Sgr dSph in the baseline model iterations}
\label{sec:skyfact-sgr-baseline}

Following the outline of our fitting rationale in Sec.~\ref{sec:skyfact}, we begin our iteration towards an optimised background model incorporating a dedicated treatment of spectral and spatial model uncertainties with an almost standard template fit as done in the previous subsection. We call this iteration \emph{Run 0}. In total, we conducted nine \sky-runs.

\noindent\textbf{Run 0.}  We employ the \sky~hyperparameter definition of Sec.~\ref{sec:skyfact-sgr-crocker-comparison} including the treatment of point-like and extended 4FGL-DR3 sources as described there. This treatment of 4FGL-DR3 sources is also kept throughout the nine \sky-fits with the baseline model. 
A summary of the employed set of hyper-parameters per model component is given in Tab.~\ref{tab:summary-baseline}. A full table with the respective definitions for all following fit iterations is provided in Tab.~\ref{tab:summary-baseline-full} of Appendix \ref{app:baseline-full-table}.

\begin{figure*}
\centering
    \includegraphics[width=0.8\linewidth]{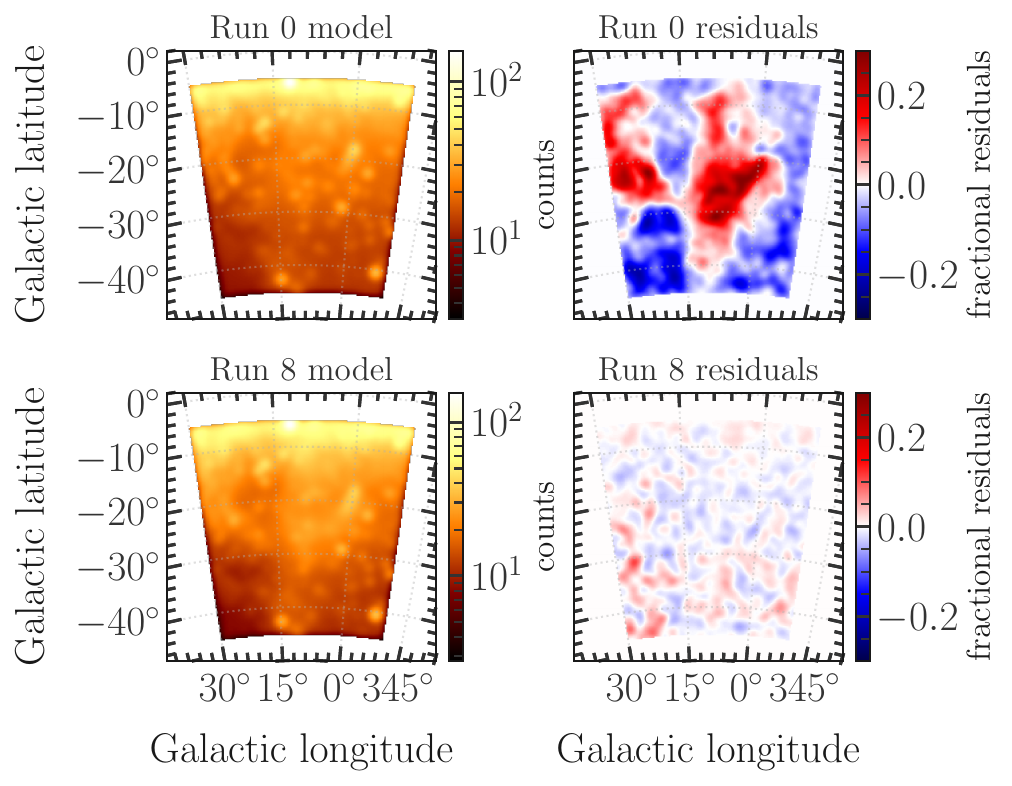}
    \caption{ 
    Comparison of the optimised model including the Sgr dSph after the \sky-run (\emph{Left}) and the resulting fractional residuals $(\phi_{\mathrm{data}} - \phi_{\mathrm{model}}) / \phi_{\mathrm{model}}$ (\emph{Right}). All displayed maps depict the integrated emission in the energy range from 1 GeV to 4 GeV. They have further been smoothed with a Gaussian kernel of $0.8^{\circ}$ roughly corresponding to the LAT PSF size at 1 GeV. The top row displays the results obtained from Run 0 with the baseline model composition as detailed in Tab.~\ref{tab:summary-baseline}, whereas the bottom row illustrates our findings in Run 8, which reflects the case with optimised background model reconstruction. Note that the ROI shown in Fig.~\ref{fig:sgr-templates} is the same as here rendering it easy to locate the position of Sgr in these plots. \label{fig:baseline-results}}
\end{figure*}

We show in the top panels of Fig.~\ref{fig:baseline-results} the best-fitting model (baseline including the Sgr dSph) and the fractional residuals $(\phi_{\mathrm{data}} - \phi_{\mathrm{model}}) / \phi_{\mathrm{model}}$ in the energy band from 1 GeV to 4 GeV. The displayed quantities are directly comparable with Figure 3 in the ``extended data'' section of \crocker. As expected, the template fit converges to a best-fitting model that leaves pronounced relative residuals up to $30\%$ of the observed gamma-ray photon counts. The residuals appear to be structured, in particular, in the central region of the ROI around the location of an arc of atomic and molecular hydrogen that also overlaps with the position of the FBs. Another striking feature is the overfitting in the lower left corner of the ROI, which arises again -- when confronted with the spatial morphology of the employed model components -- due to the gas maps. Interestingly, the same residual structure was uncovered in \crocker, which demonstrates that such a model
is not free of artefacts from mis-modelling. 

Besides the qualitative comparison between both analyses, we quantified the resulting significance of the Sgr dSph. Conducting Run 0 with a baseline model with and without the Sgr template allows us to employ Eq.~\ref{eq:significance-mixture-model}. 
We obtain strong evidence of around $14\sigma$ for the presence of the Sgr dSph in the cocoon region. We stress that at this level of the iterative runs, we employ the flat version of the FBs template, and that we can reproduce similar results to 
\crocker.

\noindent\textbf{Run 1 to Run 5.} In this sequence of \sky~runs we successively free those astrophysical background components that we expect to influence the preference in the LAT data for emission from the Sgr dSph but that are distinct enough in either spatial morphology or spectrum. 
Details of the hyper-parameters and results of subsequent runs are reported in App.~\ref{app:baseline-full-table}. In addition, we provide the summary of all fit residuals for the baseline setup in App.~\ref{app:residuals-per-run} as well as the spatial re-modulation parameters for the gas, IC and FBs (where applicable) in App.~\ref{app:modulation-per-run}.
We find that optimising the IGRB and Sun+Moon contributions does not sizeably affect the significance of the Sgr dSph. However, these two steps emphasise a critical point about the interpretation of \sky~results. As stated in Tab.~\ref{tab:summary-baseline-full}, the minimal log-likelihood value increases from Run 0 to Run 1. This does not necessarily imply that we failed to reach the global minimum. Since the total likelihood receives contributions from a Poisson likelihood and a penalising likelihood function, the increased log-likelihood value is likely due to the latter. Along these lines, we caution the reader to interpret the stated likelihood values as a qualifier of the goodness-of-fit of the currently employed gamma-ray emission model. Decreasing values are certainly indicative of a better fit to the data -- as demonstrated by the sudden jump from Run 2 to Run 3 when we start to optimise the bright diffuse components like gas-related templates (see also Fig.~\ref{fig:baseline-all-residuals} in App.~\ref{app:residuals-per-run}) -- but it is by no means a statistically robust statement. Such statements are only possible in nested model settings with the exact same hyper-parameter configuration as we do to infer the Sgr dSph's significance. 

As for residuals, when we activate spatial re-modulation of both gas-related components, we observe a sharp drop in the significance of the Sgr dSph, which is still strongly preferred by the data. The former positive residuals in the centre of our ROI are alleviated and attributed to the gas maps as are the residuals in the region of overfitting between $15^{\circ}$ to $30^{\circ}$ in longitude and $-45^{\circ}$ to $-30^{\circ}$ in latitude (cf.~with the spatial modulation of the gas templates shown in Figs.~\ref{fig:baseline-modulation-hi} and \ref{fig:baseline-modulation-h2}). To some extent, those residuals are the mere result of fitting a gas density template rather than a model of gamma-ray emission generated with cosmic-ray propagation solvers. Additionally optimising the IC component further reduces the significance of Sgr. Here, we observe that the spatial re-modulation inscribes a structure to the IC template that resembles the profile of the FBs (see the Run 5 panel of Fig.~\ref{fig:baseline-modulation-ic}). This is conceivable since we are still utilising a completely uniform FBs template and secondly, the gamma-ray production mechanisms of the IC contribution and those hypothesised for the FBs are not too different.

\noindent\textbf{Run 6.} The results of Run 5 already hinted at the importance of the FBs. The spectrum of the IC component and the FBs share similar features and are thus partially degenerate. Besides, the strong positive residuals obtained in Run 0 around the centre of the ROI overlap with the FBs and Sgr dSph region. Thus, the gas components share a certain degeneracy with the FBs. In Run 6, we inject the structured spatial model of the FBs but treat this component as in a standard template fit, i.e.~no constraint on the spectrum (see configuration in Tab.~\ref{tab:summary-baseline}). Hence, this run sheds light on the question of how much the spatial morphology of the FBs impacts the need for an additional gamma-ray contribution in its extended profile. Of course, the employed FBs template was derived in a data-driven way implying that it potentially contains emission that may stem from Sgr.

After the \sky~fit, we still find evidence for the Sgr dSph at $2.4\sigma$. However, the significance has more than halved compared to Run 5. Thus, two important factors impact Sgr's significance: The degree of mis-modelling in the Galactic diffuse emission (gas and IC maps) and the gamma-ray emission associated with the FBs. Run 6 is somewhat comparable to the ``default model'' employed in \crocker. This default model is -- paying only attention to the critical components -- comprised of their gas maps inferred from hydrodynamical simulations plus the structured FBs template of \cite{Fermi-LAT:2014sfa}. We use the same FBs characterisation and a \sky~optimised collection of gas templates. In \crocker~the significance of the Sgr dSph is quoted as $8.1\sigma$ in this setting. It is conceivable that the mis-modelling of the gas-related gamma-ray emission is partially driving the mismatch between the two results. After all, the authors of \crocker~find very similar residuals in the 1 GeV to 4 GeV energy range as we do in the upper panel of Fig.~\ref{fig:baseline-results}. The image reconstruction capabilities of \sky~help  to more reliably infer the significance of emission from Sgr. If we connect the gamma-ray emission from Sgr with a population of MSPs, then there should not be a strong spectral degeneracy between them and the gas contribution. In this sense, the optimised gas templates should not absorb much of the gamma-ray emission from the Sgr dSph. 

\noindent\textbf{Run 7.} The structured FBs template of Run 6 is the product of a study using different data selection cuts, analysis methods and gamma-ray emission models than what adopted here. Thus, we aim to derive our own characterisation of the FBs' emission in the \sky~framework. To this end, we resort again to the flat FBs template but allow for unconstrained spatial re-modulation with a rather short smoothing scale (shorter than the one applied to the gas templates). As a consequence, we fix the spectrum of the FBs to the one derived in  \cite{Fermi-LAT:2014sfa} with room for $1\%$ variations without smoothing. The overall normalisation factor is kept at a value of 1. As in the original \sky~publication, this approach avoids the unwanted loss of convexity of the optimisation target function. It also leaves enough freedom to the Sgr dSph template to absorb gamma-ray emission in the cocoon region provided that the physical mechanism for Sgr's gamma-ray emission is generating a spectrum distinct from the hard spectrum of the FBs (as would be the case in the MSP scenario).

Optimising the FBs' structure from the initial flat template further reduces the significance of emission from Sgr to $1.9\sigma$, i.e.~a sizeable upward fluctuation but no clear evidence. Evaluating the configuration of the component-specific spatial re-modulation parameters shows that we were able to break the degeneracy between the IC and FBs templates (see Fig.~\ref{fig:baseline-modulation-ic} for the IC component and Fig.~\ref{fig:baseline-modulation-fb} for the FBs). Imposing a physics-motivated FB spectrum correctly attributes the residual emission in the region from $10^{\circ}$ to $-10^{\circ}$ in longitude and $-40^{\circ}$ to $0^{\circ}$ in latitude to the FBs but not, as observed in Run 5 onwards, to the IC component. The reconstruction of the gas-related emission is only marginally affected. Therefore, we arrived at a gamma-ray emission model iteration that appears to be physically sound.

\noindent\textbf{Run 8.} This last run of the iterative fit is also the first robustness check of our findings (see configuration and results in Tab.~\ref{tab:summary-baseline}). To challenge the goodness of fit of the optimised Run 7, we lift some of the strict spectral constraints to see how degenerate the considered components really are and how fine-tuned the hyper-parameters need to be to break such degeneracies. In particular, we increase the allowed spectral variations of the gas maps from $5\%$ to $15\%$ and from $1\%$ to $5\%$ concerning the FBs. Lending more freedom to the FBs spectrum is reasonable given the fact that it also depends on the analysis details of \cite{Fermi-LAT:2014sfa}. The IC component already had considerable spectral freedom. On top of that, we slightly constrain the overall normalisation of the IGRB around the model prediction.
The results in terms of the optimised model and fractional residuals are displayed in the bottom row of Fig.~\ref{fig:baseline-results}. The character of the residuals has drastically changed compared to the standard template fit in Run 0. First and foremost, they do not seem to exhibit any structure that would hint at a missing or mis-modelled component (as expected from the scope of \sky). Most of the residuals are close to zero with a maximal positive deviation around $10\%$ in a region (lower left corner) that is not causally connected to the cocoon region. Inspecting the structure of the spatial re-modulation parameter confirms the observations made in Run 7. Our final model iteration appears to represent a valid description of reality in the selected ROI. Although certainly existent, the degeneracies between the various extended diffuse emission components are not so strong to require intricate fine-tuning of the hyper-parameters. Moderate constraints are sufficient. As a consequence, the significance of the Sgr dSph is further reduced to around $0.7\sigma$. 

\begin{figure*}
\centering
    \includegraphics[width=0.9\linewidth]{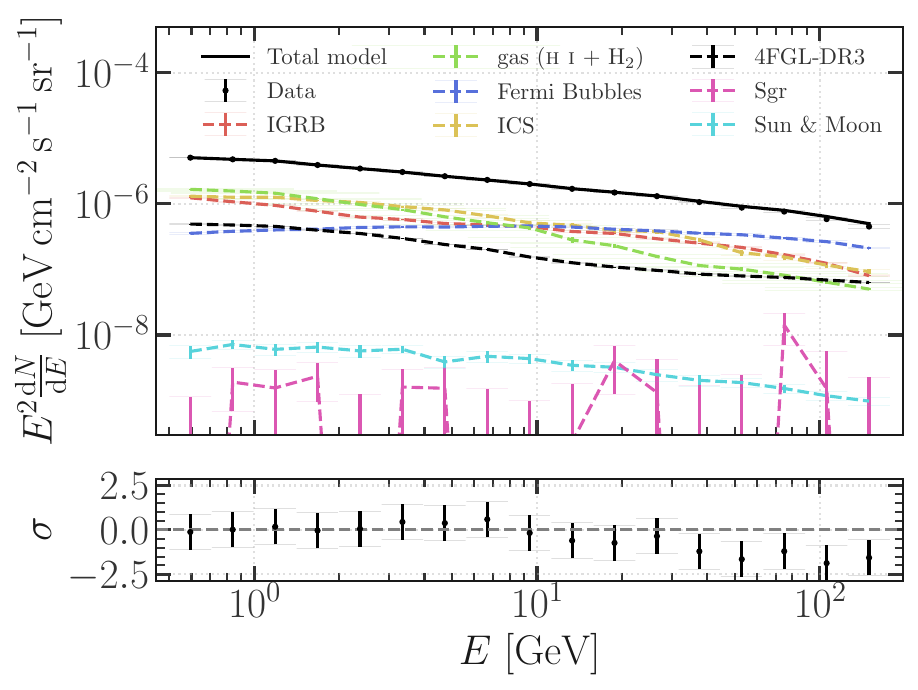}
    \caption{(\emph{Top}:) 
    Run 8 best-fitting spectrum of each baseline model component to the total gamma-ray emission within our ROI of $40^{\circ}\times40^{\circ}$ centred on $(\ell, b) = (10^{\circ},-25^{\circ})$ averaged over the corresponding angular size. We refer here to the baseline model including Sgr. The \fermi-LAT data is displayed as black data points with uncertainty matching the $1\sigma$ Poisson error. We added the inferred spectrum of the sum of all point-like and extended 4FGL-DR3 sources, which are shown in black as well. Note that the spectrum of both gas templates (green) of the baseline model appears summed as well. The error bars of the model components follow from the results of fitting 30 Poisson realisations drawn from the best-fitting model of Run 8. The indicated values refer to one standard deviation. (\emph{Bottom}:) Run 8 spectral residuals following from the \fermi-LAT data and the summed contribution of all model components. The residuals are in units of standard deviations, defined as $\sigma = (\mathrm{data}- \mathrm{model})/\sqrt{\mathrm{data}}$. \label{fig:run8-spectrum}}
\end{figure*}

We supplement the evaluation of the spatial residuals with the inferred best-fitting spectra of each component of Run 8 (incl.~the Sgr dSph) shown in the top panel of Fig.~\ref{fig:run8-spectrum}. The equivalent of this plot is displayed in Figure 7 of the extended data section of \crocker. The bottom panel of Fig.~\ref{fig:run8-spectrum} provides the spectral residuals between \fermi-LAT data and the summed baseline model in units of standard deviations. We find nearly flat residuals around zero up to $\sim30$ GeV, whereas at higher energies our best-fitting model is slightly over-fitting the data. Given the decreasing statistics at higher energies -- in particular, at higher latitudes -- the behaviour of the obtained high-energy residuals is well within the expectations. The inferred spectra of the bright diffuse and PLS contributions are broadly consistent with the results of \crocker. An intriguing difference is the profile of the FBs, IGRB and ICS components at energies above 10 GeV. We find that the FBs are the dominant contribution in this energy range -- likely due to the imposed hard spectrum measured at high latitudes -- while their spectrum in \crocker~appears to drop with increasing energy rendering the IGRB and IC components the brightest contributors. This observation may be the result of the mentioned degeneracy between IC and FBs. As discussed in the context of the spatial residuals, \sky~seems to have partially broken this degeneracy which may impact the results of \crocker. However, the slight over-fitting we observe at the highest energies can conceivably be related to the reconstructed FBs spectrum and a lack of freedom during the \sky~run. Yet, the reconstructed emission of the Sgr dSph is rather negligible and highly uncertain according to the inferred fit errors. These errors were derived by running \sky~in Run 8 configuration on 30 Poisson realisations of the best-fitting baseline model to obtain mock data of the gamma-ray sky. The error bars refer to one standard deviation around the mean. The level of the Sgr gamma-ray emission is around one order of magnitude lower than in \crocker. More importantly, we do not find a statistically robust component at energies between 1 GeV and 10 GeV; the energy regime where an MSP-like component would shine the brightest. This is consistent with the fact that the overall significance of the Sgr dSph is less than $1\sigma$. 

\begin{table*}
    \centering
    {\renewcommand{\arraystretch}{1.5}
    \begin{tabular}{l p{2cm} p{2cm} p{2cm}}
    \Xhline{5\arrayrulewidth}
    \multirow{2}{*}{Components} & Run 0 & Run 6 & Run 8\\
     & \multicolumn{3}{c}{\sky~hyper-parameters:$\bigl[\begin{smallmatrix}
        \lambda & \lambda^{\prime} & \lambda^{\prime\prime} \\
        \eta & \eta^{\prime} & \cdot 
    \end{smallmatrix}\bigr]$}\\
    \hline
    4FGL-DR3 (PLS) & $\bigl[\begin{smallmatrix}
        \cdot & 25 & 10  \\
        \cdot & 0 & \cdot
    \end{smallmatrix}\bigr]$ & $\bigl[\begin{smallmatrix}
        \cdot & 25 & 10  \\
        \cdot & 0 & \cdot
    \end{smallmatrix}\bigr]$& $\bigl[\begin{smallmatrix}
        \cdot & 25 & 10  \\
        \cdot & 0 & \cdot
    \end{smallmatrix}\bigr]$ \\
    4FGL-DR3 (ext) & $\bigl[\begin{smallmatrix}
        0 & 1 & \infty  \\
        6 & 0 & \cdot
    \end{smallmatrix}\bigr]$ & $\bigl[\begin{smallmatrix}
        0 & 1 & \infty \\
        6 & 0 & \cdot
    \end{smallmatrix}\bigr]$ & $\bigl[\begin{smallmatrix}
        0 & 1 & \infty  \\
        6 & 0 & \cdot
    \end{smallmatrix}\bigr]$\\
    gas (\textsc{h i}) & $\bigl[\begin{smallmatrix}
        \infty & 0 & 0  \\
         0 & 0 & \cdot
    \end{smallmatrix}\bigr]$ & $\bigl[\begin{smallmatrix}
        \frac{1}{25} & 400 & 0  \\
         40 & 0 & \cdot
    \end{smallmatrix}\bigr]$ & $\bigl[\begin{smallmatrix}
        \frac{1}{25} & 44 & 0  \\
         40 & 0 & \cdot
    \end{smallmatrix}\bigr]$ \\
    gas (H$_2$) & $\bigl[\begin{smallmatrix}
        \infty & 0 & 0  \\
         0 & 0 & \cdot
    \end{smallmatrix}\bigr]$ & $\bigl[\begin{smallmatrix}
        \frac{1}{25} & 400 & 0  \\
         40 & 0 & \cdot
    \end{smallmatrix}\bigr]$ & $\bigl[\begin{smallmatrix}
        \frac{1}{25} & 44 & 0  \\
         40 & 0 & \cdot
    \end{smallmatrix}\bigr]$ \\
    IC & $\bigl[\begin{smallmatrix}
        \infty & 0 & 0  \\
        0 & 0 & \cdot
    \end{smallmatrix}\bigr]$ & $\bigl[\begin{smallmatrix}
        1 & 16 & 0  \\
        150 & 0 & \cdot
    \end{smallmatrix}\bigr]$  & $\bigl[\begin{smallmatrix}
        1 & 16 & 0  \\
        150 & 0 & \cdot
    \end{smallmatrix}\bigr]$\\
    IGRB & $\bigl[\begin{smallmatrix}
        \infty & 0 & 0  \\
         0 & 0 & \cdot
    \end{smallmatrix}\bigr]$ & $\bigl[\begin{smallmatrix}
        \infty & 400 & 0  \\
         0 & 0 & \cdot
    \end{smallmatrix}\bigr]$ & $\bigl[\begin{smallmatrix}
        \infty & 400 & \frac{1}{25}  \\
        0 & 0 & \cdot
    \end{smallmatrix}\bigr]$\\
    Sun\&Moon & $\bigl[\begin{smallmatrix}
        \infty & 0 & 0  \\
         0 & 0 & \cdot
    \end{smallmatrix}\bigr]$ & $\bigl[\begin{smallmatrix}
        10 & 16 & 0  \\
         150 & 0 & \cdot
    \end{smallmatrix}\bigr]$ & $\bigl[\begin{smallmatrix}
        10 & 16 & 0  \\
         150 & 0 & \cdot
    \end{smallmatrix}\bigr]$\\
    FBs (flat) & $\bigl[\begin{smallmatrix}
        \infty & 0 & 0  \\
         0 & 0 & \cdot
    \end{smallmatrix}\bigr]$ & $\bigl[\begin{smallmatrix}
        \infty & 0 & 0  \\
         0 & 0 & \cdot
    \end{smallmatrix}\bigr]$ & $\bigl[\begin{smallmatrix}
        0 & 400 & \frac{1}{25}  \\
         6 & 0 & \cdot
    \end{smallmatrix}\bigr]$ \\
    FBs (structured) & $-$ & $\bigl[\begin{smallmatrix}
        \infty & 0 & 0  \\
         0 & 0 & \cdot
    \end{smallmatrix}\bigr]$ & $-$\\
    Sgr & $\bigl[\begin{smallmatrix}
        \infty & 0 & 0  \\
         0 & 0 & \cdot
    \end{smallmatrix}\bigr]$ & $\bigl[\begin{smallmatrix}
        \infty & 0 & 0  \\
         0 & 0 & \cdot
    \end{smallmatrix}\bigr]$ & $\bigl[\begin{smallmatrix}
        \infty & 0 & 0  \\
        0 & 0 & \cdot
    \end{smallmatrix}\bigr]$\\
    \hline
    $\mathcal{Z}_{\mathrm{Sgr}}\;\left[\sigma\right]$ & $13.6$ & $2.4$ & $0.7$\\
    \Xhline{5\arrayrulewidth}
    \end{tabular}}
    \caption{Selected results of the iterative \sky-runs based on the baseline setup outlined in Sec.~\ref{sec:sgr_physics} and the selected \sky~hyper-parameters for each model component. The first row of hyper-parameters controls the allowed variation in the magnitude of the re-modulation parameters affecting spatial pixels $\bm{\tau} \longleftrightarrow \lambda$, the spectrum per energy bin $\bm{\sigma} \longleftrightarrow \lambda^{\prime}$ and the overall normalisation $\nu \longleftrightarrow \lambda^{\prime\prime}$. The second row lists hyper-parameters determining the employed smoothing of the spatial and spectral re-modulation parameters $\bm{\tau} \longleftrightarrow \eta$ and $\bm{\sigma} \longleftrightarrow \eta^{\prime}$. The magnitude hyper-parameters are fixed to unity whenever their value is quoted as $\infty$ while they are completely free to vary when set to $0$. It holds that the larger the hyper-parameter's value, the more constrained the re-modulation parameter to fluctuate around $1$. In contrast, a smoothing scale with a value of $0$ means that no smoothing is applied. The larger the value of the smoothing hyper-parameter, the stronger the smoothing or in other words, the longer the considered correlation length among pixels or energy bins. The significance of adding a Sgr dSph component according to Eq.~\ref{eq:significance-mixture-model} is reported as $\mathcal{Z}_{\mathrm{Sgr}}$ in standard deviations.  
    \label{tab:summary-baseline}}
\end{table*}

We stress that, in contrast to traditional template fitting, where template normalisation parameters are fitted independently for each energy bin, our use of \sky~allows for correlated energy bins by incorporating physics-informed spectral priors. This approach leverages more information than standard template fits. While the FBs template from \crocker~is entirely data-driven and dependent on specific data selections, we derive our own data-driven FBs template through \sky, which hence is self-consistent with the other model components and aligned with the selected \textit{Fermi}-LAT dataset in the Sgr ROI. We also verified that the scatter of the spatial re-modulation parameters regarding the bright diffuse components like gas, IC and the FBs is reasonably distributed around the spatial prior indicating no particularly worrisome overfitting.

\subsection{Alternative model components and other model systematic uncertainties}
\label{sec:baseline-robustness}
The iterative fit rationale utilising the baseline model revealed no significant emission associated with the Sgr dSph. While we allowed for a generous amount of re-modulation of the initial templates of the baseline setup, we examine in this section the impact of the initial model component morphology on the significance of Sgr. We make use of the alternative set of templates detailed in Sec.~\ref{sec:sgr_physics} to scrutinise the fit results of the \sky~runs summarised in Tabs.~\ref{tab:summary-baseline} and \ref{tab:summary-baseline-full}. 

We do not perform the full chain of systematic runs for each alternative model composition. Instead, we explore the robustness of certain aspects of the baseline analysis when different assumptions are made. The considered robustness checks with alternative model compositions are summarised in Tab.~\ref{tab:summary-alternative}. In what follows, we provide details of the intention for each of these runs and interpret the obtained findings.

\begin{table*}
    \centering
    {\renewcommand{\arraystretch}{1.5}
    \begin{tabular}{lcccccccccc}
    \Xhline{5\arrayrulewidth}
    \backslashbox{Model}{Run}& Run 0 & Run 1 & Run 2 & Run 3 & Run 4 & Run 5 & Run 6 & Run 7 & Run 8\\
    \hline
    SgrGaia3 & $\times$ & $\times$ & $\times$ & $\times$ & $\times$ & $\times$ & $\checkmark$ & $\checkmark$ & $\checkmark$\\
    Cocoon & $\times$ & $\times$ & $\times$ & $\times$ & $\times$ & $\checkmark$ & $\times$ & $\times$ & $\times$\\   
    FB2017 & $\times$ & $\times$ & $\times$ & $\times$ & $\times$ & $\times$ & $\checkmark$ & $\times$ & $\times$\\ 
    ModelA & $\times$ & $\times$ & $\checkmark$ & $\checkmark$ & $\checkmark$ & $\checkmark$ & $\checkmark$ & $\checkmark$ & $\checkmark$\\
    Model0 & $\checkmark$ & $\times$ & $\times$ & $\times$ & $\times$ & $\times$ & $(\checkmark)$ & $\times$ & $(\checkmark)$\\ 
    \Xhline{5\arrayrulewidth}
    \end{tabular}}
    \caption{Summary of the robustness checks performed with alternative gamma-ray emission models as outlined in Sec.~\ref{sec:sgr_physics} and the beginning of Sec.~\ref{sec:baseline-robustness}. We indicate the type of performed run per model instance which is based on the hyper-parameter definitions shown in Tab.~\ref{tab:summary-baseline}, and Tab.~\ref{tab:summary-baseline-full}. Components exchanged in comparison to the baseline model are initialised with hyper-parameters corresponding to their equivalent in the baseline setup. The parentheses in the last row indicate a treatment of model components that differs from the standard approach. Details are given in the text.
    \label{tab:summary-alternative}}
\end{table*}

\paragraph*{Significance of the cocoon region.} We found for the baseline setup that Run 6 decreases the significance of the Sgr dSph component to $2.4\sigma$. The difference between Run 5 and 6 is the change from a featureless FBs template to a structured template including the cocoon region's emission. The observed drop in significance is natural if Sgr's gamma-ray emission is causing the enhanced brightness of the cocoon, namely if the data-driven FBs template absorbs 
the Sgr emission in the brightening of the cocoon region. Therefore, we explore how significant the cocoon itself is compared to a fully uniform emission of the FBs. 
We adopt the \sky~hyper-parameters of Run 5 to perform this cross-check thereby guaranteeing optimised diffuse background components. We note that this setup is simply exchanging the Sgr dSph template in Tab.~\ref{tab:summary-baseline} with the cocoon template.

The resulting minimal log-likelihood value is $-2\ln{\mathcal{L}}_{\mathrm{base+cocoon}} = 297183$ implying a significance of the cocoon region of $\mathcal{Z}_{\mathrm{Cocoon}} = 20.3\sigma$. Despite the cocoon's significance being higher than the corresponding significance of Sgr for Run 5, we cannot conclude that the cocoon's morphology is a better model for this region of the sky. 
In a frequentist approach, we consider a nested model containing both, the Sgr dSph and cocoon template. This results in a likelihood value of $-2\ln{\mathcal{L}}_{\mathrm{base+cocoon+Sgr}} = 297159$, which means that there is 2.7$\sigma$ evidence that an additional Sgr dSph component on top of the cocoon region is preferred by the data. Conversely, we may start with a model including backgrounds and Sgr and add the cocoon in the second, nested model step. We already tested the first composition in our systematic baseline runs, the likelihood is stated in Tab.~\ref{tab:summary-baseline-full} reading $-2\ln{\mathcal{L}}_{\mathrm{base+Sgr}} = 297584$. From this, the significance of the additional cocoon component is $\mathcal{Z}_{\mathrm{cocoon}} = 19.3\sigma$, i.e.~very strong evidence and only slightly reduced in comparison with the background-only fit.
In other words, the additional Sgr component does not yield a substantial improvement of the fit and it is, hence, not required given the \textit{Fermi}-LAT data. We note that the obtained significance above is just slightly larger than the result we derived in Run 6 with the complete structured FBs template. Both approaches are consistent with each other. 

One may wonder why the value of $-2\ln{\mathcal{L}}_{\mathrm{base+cocoon}}$ is much lower than $-2\ln{\mathcal{L}}_{\mathrm{base}}$ for Run 6. At first glance, the difference between the two model instances is the added spatial structure of the FBs in all of their extent in the sky. However, this is not completely true since this specific Run 5 definition allows the cocoon region to attain a spectrum and total luminosity independent of the rest of the FBs while the single structured FBs template in Run 6 keeps this relative normalisation fixed. Run 5 with the cocoon is, hence, the more flexible model, which may easily explain a substantial difference in the likelihood function's value. Interestingly, the Run 5 reconstructed spectrum of the FBs and the cocoon are almost identical (both are initialised as flat power laws) while the FBs' gamma-ray flux is slightly higher than the one of the cocoon region. 

\paragraph*{Changing the spatial profile of the Sgr dSph.} We consider the SgrGaia3 model (see Sec.~\ref{sec:sgr_physics} and Fig.~\ref{fig:sgr-templates} for details). 
Performing the analyses for each run, we obtain the following likelihood and significance values:
\begin{itemize}
    \item Run 6: $-2\ln{\mathcal{L}}_{\mathrm{base+SgrGaia3}} = 297405$ and $\mathcal{Z}_{\mathrm{SgrGaia3}} = 3.7\sigma$,
    \item Run 7: $-2\ln{\mathcal{L}}_{\mathrm{base+SgrGaia3}} = 296209$ and $\mathcal{Z}_{\mathrm{SgrGaia3}} = 1.6\sigma$,
     \item Run 8: $-2\ln{\mathcal{L}}_{\mathrm{base+SgrGaia3}} = 296139$ and $\mathcal{Z}_{\mathrm{SgrGaia3}} = 0.5\sigma$.
\end{itemize}
We do not find substantial differences compared to the baseline setup with a potential exception of Run 6, which yields larger evidence for the Sgr dSph component almost reaching $4\sigma$. However, this pronounced significance does not propagate to the next runs when we re-modulate the FBs with \sky. The more substantial significance may arise due to the altered spatial profile of Sgr. The \textit{Gaia} Data Release 3 iteration is less peaked towards the core of Sgr and traces larger parts of the cocoon region. Hence, it may simply provide a suitable description of the cocoon in addition to the emission modelled via the structured FBs template. 


\paragraph*{Impact of the FBs spatial template.} We assessed in Run 6 Sgr's significance via a structured FBs template that had been derived in a data-driven fashion. The corresponding original publication neither used the data selection adopted in this work nor the model compositions or statistical framework. Accordingly, the obtained spatial morphology of the FBs is subject to many assumptions. To understand the impact of such external assumptions in our analysis, we now investigate the FB2017 model with the same techniques as the model in our baseline setup. Employing the same Run 6 hyper-parameters, we ask the question of Sgr's significance. 

We obtain $-2\ln{\mathcal{L}}_{\mathrm{base}} = 297031$ and $-2\ln{\mathcal{L}}_{\mathrm{base+Sgr}} = 297000$, which corresponds to a significance of $\mathcal{Z}_{\mathrm{Sgr}} = 3.4\sigma$. Compared to the baseline setup's significance this is a vast jump of $1\sigma$ yielding a \textit{hint} for emission from Sgr. The most likely explanation for this difference is the luminosity of the cocoon region in the FBs template of \cite{Fermi-LAT:2017opo}. It is on average 25\% less bright than the version utilised in the baseline setup, while in extreme cases the difference can be as large as 45\%. The data seem to require a larger cocoon flux than this model provides such that the Sgr dSph template fills the gap rendering it more significant. 
However, we demonstrated before that a flat cocoon is preferred more than 
the spatially structured Sgr template.

These findings also emphasise the advantage of our adaptive-template fitting approach as we can self-consistently derive a data-driven FBs spatial morphology from the selected \textit{Fermi}-LAT data.

\paragraph*{Impact of the assumed gas density in the Milky Way.} The systematic exploration of the adaptive template fit regarding the baseline setup relied on a particular set of spatial templates for the $\pi^0$ (+ bremsstrahlung) component of the Milky-Way diffuse foreground emission. Its structure is directly related to the distribution of atomic and molecular gas in our Galaxy. It was repeatedly stressed in \cite{Song:2024iup} that \sky~is not a perfect fitting routine as changes in the spatial morphology of the initial background templates may eventually result in different optimised background templates, i.e.~there is no unique solution to how to re-modulate the input. Therefore, we consider the gas component of FGMA as an alternative to our baseline choice. A crucial difference between both models is the achieved resolution of the underlying gas column density maps of the Galaxy. FGMA is based on the works of \cite{2001ApJ...547..792D, 2005A&A...440..775K} with an angular resolution of about $36^{\prime}$ (full width at half maximum) \cite{2016A&A...594A.116H} compared to $16\overset{\prime}{.}2$ of the \textsc{hi4pi} survey used as the input of our baseline setup. Moreover, FGMA yields the gamma-ray flux from the summed atomic and molecular hydrogen contributions calculated via the numerical cosmic-ray propagation solver \texttt{GALPROP}. These differences render it possible to study the robustness of our baseline results regarding the priors on the Milky-Way gas distribution, which we will scan for runs 2 to 8. We prepare a summary of our findings in Appendix \ref{app:modelA-runs} displayed in Tab.~\ref{tab:summary-alternative-modelA} in full analogy to the baseline setup (except for Run 3 and 4, which are the same due to a single gas template). 

We do not observe substantial differences between ModelA and our baseline setup regarding the evolution of Sgr's significance per fitting run. The employed Sgr dSph component is significant when no structure within the FBs is assumed but drops below $3\sigma$ once we accounted for it. The numerical values for the respective significances are fairly consistent with the baseline results. Therefore, we are inclined to conclude that the initially assumed gas distribution does not have a strong impact on our analysis results. 

\paragraph*{Utilising the gamma-ray emission model of \protect\crocker.} As an internal cross-check, we perform a template-based fit (Run 0 hyper-parameters) based on the benchmark model of \crocker~to reproduce their results as best as possible within the \sky~framework (Model0). We obtain likelihood values of $-2\ln{\mathcal{L}}_{\mathrm{Model0}} = 308644$ and $-2\ln{\mathcal{L}}_{\mathrm{Model0+Sgr}} = 308234$ corresponding to a significance of the Sgr dSph template of $\mathcal{Z}_{\mathrm{Sgr}} = 18.9\sigma$. Such high significance was also obtained in \crocker~as reported in their Table 1 for a flat (or uniform) FBs spatial template.

While it would be desirable to investigate the full evolution from Run 0 to Run 8 with the original model of \crocker, it appears to be impossible for all practical matters. Its diffuse components are split into rings while they were individually normalised, which prevents us from adding them together without a prior fit to the data. Therefore, we would have to keep all components in the fit and re-modulate each of them. In this situation, we would introduce potentially large degeneracies between the components as they are not spatially distinct enough. Thus, the \sky~fit is doomed to fail due to the non-convexity of the minimisation problem. 

As an alternative, we proceed exactly as we say: We use the fit results of Run 0 to fix the relative normalisation parameters for all ring-based components to generate a single gas and IC template, respectively. The obtained spatial priors serve as proxies for the full analysis in the setting of Run 6 and Run 8. We derive the following likelihood values and significance values for the Sgr dSph:
\begin{itemize}
    \item Run 6: $-2\ln{\mathcal{L}}_{\mathrm{Model0}} = 297522$ and $-2\ln{\mathcal{L}}_{\mathrm{Model0+Sgr}} = 297480$ corresponding to a significance of the Sgr dSph template of $\mathcal{Z}_{\mathrm{Sgr}} = 4.4\sigma$,
    \item Run 8: $-2\ln{\mathcal{L}}_{\mathrm{Model0}} = 296247$ and $-2\ln{\mathcal{L}}_{\mathrm{Model0+Sgr}} = 296233$ corresponding to a significance of the Sgr dSph template of $\mathcal{Z}_{\mathrm{Sgr}} = 1.4\sigma$.
\end{itemize}
While the results for Run 8 are in line with those obtained for the baseline and alternative compositions, Run 6 detects Sgr with a rather high significance above $4\sigma$. However, we stress that this result should not be over-interpreted because of the way we derived the gas and IC templates. They stem from Run 0, which we have shown to yield rather large residuals; the Model0 composition is no exception. Since \sky~cannot re-modulate any unphysical relative normalisation of the ring-based templates that might occur due to Run 0, the results for Run 6 may be driven by remaining background mis-modelling. In general, this should also affect Runs 7 and 8, but we expect such artefacts to be removed because of the greater freedom in the FBs. 

\subsection{Sagittarius stellar stream}
\label{sec:stream}

The complete Sgr dSph density map from \textit{Gaia} DR3 shown in Fig.~\ref{fig:rois} reveals two large-scale streams extending in the southern and northern hemispheres that could, in principle, be hosting a faint gamma-ray emission correlated with the putative population in Sgr's core. 
While the baseline template is explicitly meant to model the core region and does not contain any significant stream component, the alternative model of the Sgr dSph density map derived from the \textit{Gaia} DR3 selection is comprised of stars within Sgr's core and the larger stellar stream.   
However, to build this density model, an empirical threshold of 18 stars is applied. 
To test for a possible preference for gamma-ray emission from the full Sgr stellar stream, we repeated the \sky~analysis in the Sgr ROI with Run 6 hyper-parameters using a template derived without applying any stellar threshold. However, accounting for this extended morphology does not significantly improve the fit. Quantitatively, we find a likelihood value of $-2\ln{\mathcal{L}}_{\mathrm{base+SgrStream}} = 297399$ corresponding to a significance of the full stream component of $\mathcal{Z}_{\mathrm{SgrStream}} = 4.2\sigma$ over the background. Yet, the additional significance of the stream on top of the core region of Sgr is only $\mathcal{Z}_{\mathrm{SgrDwarf+SgrStream}} = 0.4\sigma$. The significance values drop noticeably for Run 7 and 8 fully in line with the reported results for the alternative Sgr template. 
From this observation, we expect that performing a similar template-based analysis in the northern part of the Sgr stellar stream (see the density in Fig.~\ref{fig:rois}) will not yield a strong indication of the necessity of such a component.
In fact, if compared to the stellar stream in the northern hemisphere, a similar stellar density is found in the southern part of the stellar stream included in our southern ROI.
While a full sky analysis of the correlated northern and southern stream could in principle be informing, we refrain from embarking in such a fit given the expected diffuse mis-modelling in large fitting regions in the sky, which could not be efficiently addressed even using \sky~framework, given the dimension of the parameter space involved. In addition, a combined analysis of the northern and southern gamma-ray sky requires a suite of dedicated simulations to optimise the analysed ROI regarding the statistical expectations of, e.g., the scatter of the employed test statistic. Otherwise, the derived findings are prone to becoming overconfident (see \cite{Eckner:2022swf}).

\section{Dissecting the cocoon region with pixel-count statistics}
\label{sec:results-1ppdf}
By using the 1pPDF on different ROIs, and using gamma-ray diffuse emission models as optimised with \sky, we aim at complementing the results reported in the previous section by searching for non-Poissonian components of the detected gamma rays compatible with a point-source origin of the Sgr dSph emission. In addition, using the Bayesian framework of the 1pPDF, fits to the gamma-ray sky using different diffuse emission models (e.g. including or not the \sky~ re-modulations, the FB structures, the Sgr dSph...) are compared, while accounting for unresolved sources. 

We consider the diffuse emission models as resulting from the \sky~ Run~0, Run~6 and Run~8, and repeat the fit with and without including the Sgr dSph template for each model. This is done to test if the emission attributed to a diffuse Sgr dSph template within \sky~is degenerate or better explained with point-like contribution to the $dN/dS$ within the ROI. 

When self-consistently including in the 1pPDF fit the Sgr dSph template obtained within the \sky~ fit, the Bayesian log evidence values  
$\ln(\mathcal{Z})$ are found to be -6139, -6098, -6083) for Run~0, Run~6 and Run~8 models respectively. 
When running the 1pPDF only with the diffuse emission model template, and without the Sgr dSph template used to optimise the fit to the gamma-ray sky within \sky, we obtain 
$\ln(\mathcal{Z})=$-6144, -6099, -6084, again for the Run~0, Run~6 and Run~8 models respectively. 
We thus find that the model that describes better the \textit{Fermi}-LAT data within the 1pPDF fit is the one resulting from the \sky~Run~8, for which the Sgr dSph emission is found to be not significant. Consistently, within the 1pPDF fit, we encounter similar $\ln(\mathcal{Z})$ when including or not the emission (not statistically significant) attributed to the Sgr dSph. 
We observe that Run~8 and Run~6 models are always preferred to Run~0 models and that adding or not the Sgr dSph template does not strongly affect the results. 
Comparing Run~0 including the Sgr dSph template to Run~8 without including it, we note that improving the diffuse model with \sky~has a much bigger impact on the description of the gamma rays in the ROI within the 1pPDF fit compared to the addition of the Sgr dSph emission. 

\begin{figure*}
\centering\includegraphics[width=0.49\linewidth]{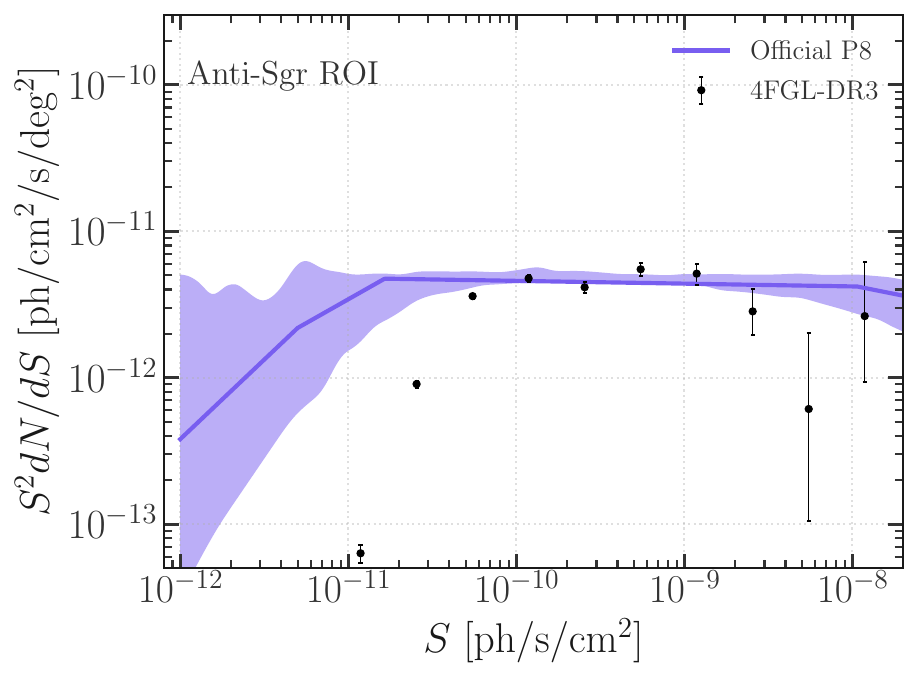}
\includegraphics[width=0.49\linewidth]{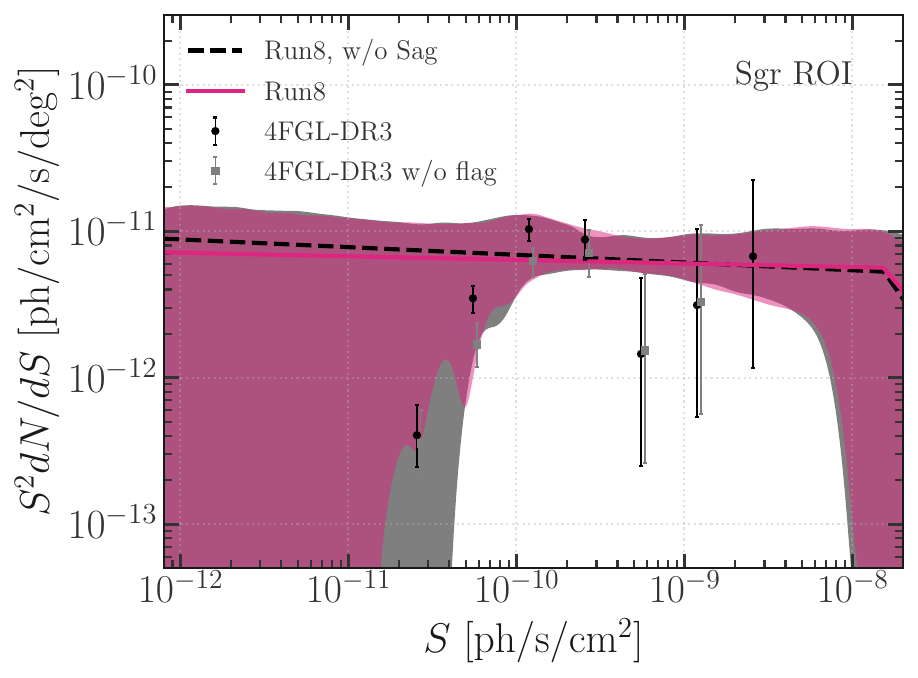}
    \caption{Source-count distribution of the Anti-Sgr ROI (left panel) and Sgr ROI (right panel) as measured by the 1pPDF in the 2--5~GeV energy range. 
The $1\sigma$ uncertainty bands are indicated with colored areas.
For the Sgr ROI results, the diffuse emission model templates from Run~8 as optimized with \sky~are used. 
The count distribution of 4FGL-DR3 sources is illustrated with black points for comparison in the bright flux regime. Gray points in the right panel represent the count distribution of 4FGL-DR3 sources without any analysis flag, see \cite{Fermi-LAT:2019yla}. \label{fig:1ppdf_realsky}}
\end{figure*}

The 1pPDF measures the $dN/dS$ as averaged in a given ROI, assuming a single, cumulative population of point sources. At the Galactic latitudes of the Sgr dSph, the bright and faint point sources are expected to be a mixture between the Galactic disc and extragalactic populations. 
Relying on a strategy introduced in \cite{Calore:2021jvg,Manconi:2024tgh}, to investigate if within the Sgr dSph ROI (solid white ROI in Fig.~\ref{fig:rois}) an additional population of faint sources is present, we measure and compare the $dN/dS$ in a control region covering the same latitudes, but complementary in longitude, the Anti-Sgr region as defined in Fig.~\ref{fig:rois}. 
The results for the $dN/dS$ measured in the Anti-Sgr (Sgr) ROI are reported in the left (right) panel of Fig.~\ref{fig:1ppdf_realsky}.
The lines illustrate the 1pPDF best fit for the source-count distribution within the ROI, and the shaded bands the $1\sigma$ uncertainties. The results are compared with the $dN/dS$ of resolved sources as found in the 4FGL-DR3. 
In the Anti-Sgr ROI, the measurement delivered by the 1pPDF when using the Official Pass8 diffuse template is consistent with the 4FGL-DR3 source counts up to $10^{-10}$ ph/s/cm$^2$, and extends the knowledge of the $dN/dS$ down to about one order of magnitude lower fluxes. 
As for the Sgr ROI, given the lower photon statistics included in the smaller region, the 1pPDF result is again found to be consistent with the catalogue counts, but the reconstruction at fluxes lower than $10^{-10}$ ph/s/cm$^2$ is hampered by large uncertainties. This can be understood by looking at the available photon counts for the two regions, which consist of 22371 counts in the Sgr ROI, and 350851 in the Anti-Sag ROI, more than a factor of 10 difference.
While we report the results using the benchmark diffuse emission models as obtained within \sky~Run~8, equivalent outcomes are found for Run~6. 
As for Run~0, in this case, the 1pPDF measures a slightly higher $dN/dS$ in the low flux regime; this is likely to be interpreted as diffuse mis-modelling effect \cite{Calore:2021jvg,Manconi:2024tgh}, even if the results are overall comparable at $1\sigma$ with the ones of Run~8.
Nevertheless, we do not observe a significant difference in the $dN/dS$ in the Sgr ROI with respect to the Anti-Sgr ROI, and thus no evidence for the need for additional source populations concerning the disk/extragalactic ones.
Following the $\ln(\mathcal{Z})$ results, also the $dN/dS$ measurement is not influenced by the inclusion or not of the Sgr template in the fit (dashed line in the right panel of Fig.~\ref{fig:1ppdf_realsky}). 
We stress that the diffuse emission models used to reconstruct the $dN/dS$ in the Sgr and Anti-Sgr ROI are different, as for the second we use the Official P8 model. We have explicitly verified that the results for the $dN/dS$ when using the Official P8 model in the Sgr ROI are compatible with the ones shown in Fig.~\ref{fig:1ppdf_realsky}. However,   the results obtained at low Galactic longitudes within the 1pPDF are more robust when using the \sky-optimized diffuse models, to have consistent models for the Fermi Bubbles, and the putative emission from the Sgr dwarf.

\section{Could we have missed a signal?}
\label{sec:simulations}
In light of the above results, we ask the question of how luminous an MSP population in the Sgr dSphs has to be to produce a significant signal detectable, i.e.~could we have missed a signal given our analysis setups?

In what follows, we outline the model and data generation for two scenarios of the MSP population in Sgr, and discuss the results obtained with \sky~when searching for such a signal in mock data.

\subsection{Simulations of the putative MSP population}
\label{sec:sgr-msp-sim}
The discussion reported in \crocker~ suggests that the detected gamma-ray emission from the Sgr dSph is produced by about $N_{\rm Sgr}\sim 650$ MSPs. This is estimated on the basis of their stellar population synthesis model, based on the accretion-induced collapse (AIC) MSP formation mechanism \cite{Gautam:2021wqn}. 
This population, emitting a total gamma-ray flux of $2 \cdot 10^{-8}$ ph/cm$^2$/s (as characterised in \crocker~ in  the 0.5--150~GeV), is expected to be associated with a total gamma-ray luminosity of $L_{\gamma \rm tot, Sgr}=3.8 \cdot 10^{36}$~erg/s. 
Since a robust detection is not achieved in the present work, we here present some simple benchmark simulation scenarios. 

\paragraph*{Case 1.} In this first scenario, the flux from the putative Sgr MSPs is simulated to be smoothly distributed, following the baseline Sgr template introduced in Sec.~\ref{sec:sgr_physics} and the middle panel of Fig.~\ref{fig:sgr-templates}. 
This case does not require any luminosity function modelling and assumes that the emission from the Sgr MSPs cannot be disentangled from a truly diffuse flux.

\paragraph*{Case 2.} More likely, the MSPs in the Sgr dSph follow a specific luminosity function translating into a potentially observable flux distribution. 
A luminosity function model is needed to estimate the total number of MSPs residing in the Sgr dSph and their flux distribution. We base our modelling 
on the \crocker~interpretation of MSP formation through the AIC mechanism.
In \textit{Case~2} we simulate the Sgr dSph MSPs having AIC-inspired fluxes and a spectral energy distribution modelled as a power law with index $-2.1$ in the 0.5--150~GeV interval, namely the best-fit power-law index found for the Sgr dSph spectrum in \crocker.  Since the parametrisation of the luminosity function corresponding to the AIC MSP formation mechanism within the Sgr dSph is not given in \crocker, we rely on the following strategy to simulate the MSPs fluxes. 
We start from the AIC luminosity function as provided for the Milky Way's stellar bulge \cite{Gautam:2021wqn,Dinsmore:2021nip}. We then assume that the total luminosity of each MSP population (Milky-Way bulge and Sgr) is roughly $L_{\rm tot}=N\cdot L_{\rm mean}$, where $N$ is the total number of MSP and $L_{\rm mean}$ the mean luminosity. 
By using the $N$ and $L_{\rm tot}$ as estimated in \crocker~for Sgr (650, $L_{\gamma \rm tot, Sgr}=3.8 \cdot 10^{36}$~erg/s), and the $N$ and $L_{\rm tot}$ corresponding to the Milky-Way bulge \cite{Gautam:2021wqn,Bartels:2017vsx} ($N_{\rm bulge}\sim 3 \cdot 10^5$ and $L_{\gamma \rm tot, bulge}=1.7 \cdot 10^{37}$~erg/s.),
we derive a rescaling factor for the mean luminosity  of the Sgr MSPs from the ratio $L_{\gamma \rm tot, Sgr}/L_{\gamma \rm tot, bulge}$. 
The rescaling factor (of the order of $\sim 100$) is used to rescale the luminosity of each MSP extracted from the available bulge AIC luminosity function. 
We verified \textit{a posteriori} that by simulating 650 MSPs following the AIC luminosity function of the Galactic bulge, and rescaling by the factor obtained above, we consistently obtain a total luminosity and flux in the uncertainty intervals provided in \crocker. Simulations not falling in this range are discarded. 
To translate the simulated luminosities to fluxes, a distance of $26.5$~kpc is assumed for each MSP. 
Finally, longitude and latitude are assigned to each MSP by following a probability distribution function obtained from the spatial template of the baseline Sgr model (see Tab.~\ref{tab:model-components}). 

\begin{figure*}
\centering\includegraphics[width=0.49\linewidth]{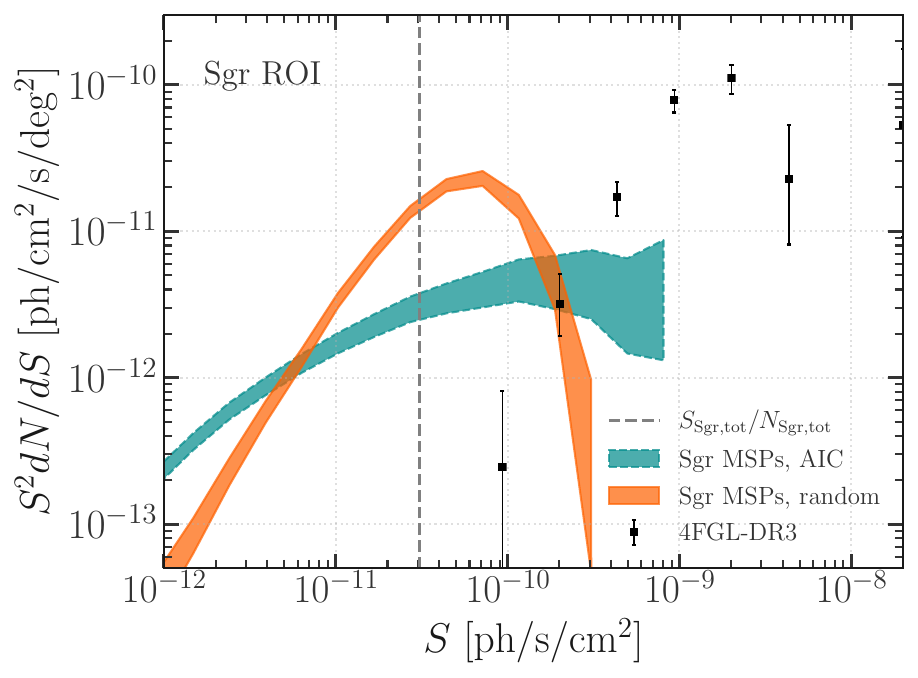}
\includegraphics[width=0.49\linewidth]{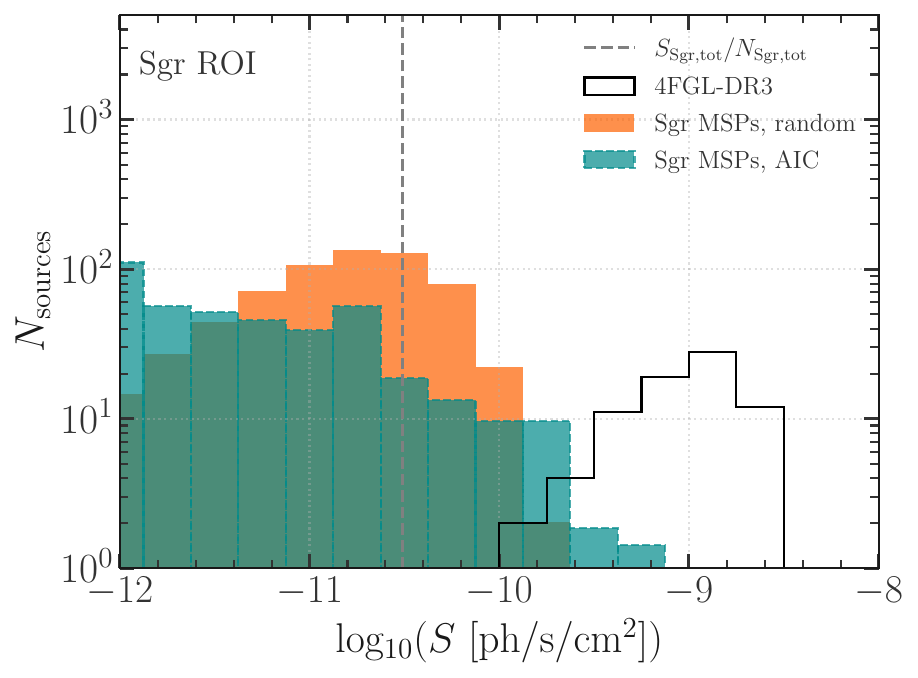}
\caption{
Simulated source-count distribution for the putative population of 650 MSPs in the Sgr dSph for gamma-ray fluxes in the 0.5--150~GeV energy range.  The cyan band (dotted contours) is obtained assuming the AIC luminosity function from \cite{Gautam:2021wqn} rescaled as described in the text, while the orange band assumes a random luminosity distribution. The dashed grey line indicates the average flux value of the MSPs population assuming all sources have the same flux and a total flux as in \protect\crocker. 
The 4FGL-DR3 sources are illustrated with black points for comparison. The left panel shows the $S^2 dN/dS$ as a function of the flux $S$, while the right panel the histogram of the source number in logarithmic $S$ bins.
\label{fig:sag_sim}}
\end{figure*}

In Fig.~\ref{fig:sag_sim} (left panel) we illustrate the results for the simulation of the Sgr MSPs following the AIC luminosity function. The expected  $dN/dS$ in the Sgr ROI for the 0.5--150~GeV energy bin is reported in cyan, and is compared to the source-count distribution of the 4FGL-DR3 detected sources within the same ROI. 
The band is obtained by computing the Poisson errors from 50 realisations of the AIC population.
The simulated MSP population is expected to be very dim and to contribute to the $dN/dS$ in the ROI in the faint-source regime ($S<10^{-9}$ ph/cm$^2$/s).
Extrapolating the $dN/dS$ of the Galactic disc and extragalactic sources to be likely a power law with an index of about $2$ in the faint regime (cf.~Fig.~\ref{fig:1ppdf_realsky}) up to  $10^{-11}$ ph/cm$^2$/s, the population of Sgr MSPs is expected to contribute less than 1\% to the observed $dN/dS$ in this regime, and thus it would be very challenging to detect or disentangle this population from other components by inspecting the cumulative source-count distribution within the Sgr ROI.  

To demonstrate that this conclusion is not affected by our approximate rescaling of the AIC luminosity function for the Sgr dSph, two other examples are investigated. 
Assuming no knowledge of the underlying MSP luminosity function, if we simulate 650 random fluxes that add up to the total, Sgr gamma-ray flux in the 0.5--150~GeV of $2 \cdot 10^{-8}$ ph/cm$^2$/s we obtain the source-count distribution reported with an orange band in the left panel of Fig.~\ref{fig:sag_sim}.  
Also in this case, the $dN/dS$ is peaked at lower fluxes with respect to the bulk of 4FGL-DR3 detected sources and can contribute at most about $10$\% of the $dN/dS$ in the faint source regime.  
Finally, we note that by just dividing the total gamma-ray flux claimed by \crocker~ in the 0.5--150~GeV energy range by $N_{\rm Sgr,tot}$ we obtain $\sim3 \cdot 10^{-11}$ ph/cm$^2$/s, which lies at least one order of magnitude lower than the detection threshold of the 4FGL-DR3 catalogue, as clear when comparing the black points in Fig.~\ref{fig:sag_sim} (left panel) with the dashed grey line. 

A complementary view of the source-count distribution for the different simulations inspected above is given in the right panel of Fig.~\ref{fig:sag_sim}, where the histograms of the number of sources as a function of their flux are given.
We observe that the MSPs from the Sgr dSph are expected to be numerous at 1-2 orders of magnitude lower fluxes with respect to the peak of sources currently detected in the 4FGL-DR3 within the ROI. 

\subsection{Mock data generation and analysis}
We resort to the \textit{Fermi Science Tools} to simulate the different MSP population cases discussed in the previous section. The Sgr MSP scenarios will be part of synthetic mock datasets that also feature the astrophysical background contributions listed in Tab.~\ref{tab:model-components} of Sec.~\ref{sec:sgr_physics}. In detail, each mock data realisation is comprised of the following components:
(i) 4FGL-DR3 point-like and extended sources,
(ii) gas from FGMA,
(iii) IC from FGMA,
(iii) structured FBs,
(iv) IGRB and
(v) Sun \& Moon.
The templates are injected with their nominal total flux derived and defined in the respective references. For the different MSP population cases, we always adopt a power-law spectrum following $N_0 \left[E/(10^{3}\;\mathrm{MeV})\right]^{-2.1}$. In {\it Case 2}, the spectral normalisation $N_0$ for each respective MSP is generated during the population creation in order to reproduce the observed total flux in \crocker. In {\it Case 1}, the fully diffuse MSP population, we set $N_0$ to achieve a certain total flux from 500 MeV to 150 GeV when integrating over the angular size of the baseline Sgr template. All background and signal components are simulated with the  \textit{Fermi Science Tools} routine \texttt{gtmodel} using the spatial and spectral binning defined for the LAT data in Tab.~\ref{tab:data_selections} (\sky~analysis). \texttt{gtmodel} generates so-called Asimov datasets, i.e.~the mean expected counts given an infinite number of Poisson realisations of the same model. Therefore, we can draw Poisson samples from the generated mock data to obtain independent model realisations for statistical tests.

The \sky~analysis of MSP {\it Case 1} and {\it Case 2} each entails different objectives to probe distinct claims made in \crocker~about the putative gamma-ray emission of Sgr, namely:
\begin{itemize}
    \item[\textit{Case 1}:] 
    The primary purpose of this case is to determine whether the \sky~method is even capable of detecting a signal from Sgr. As stated in the previous section, we conduct this check via a signal injection and recovery test. In spirit, our approach is very similar to the injection tests performed in \crocker~whose results are presented in their ``extended data'' figure 4. 
    We conduct a signal injection test to recover the injected total flux of the Sgr component. To this end, we stay within the hyper-parameter setting of Run 6 and prepare mock datasets with a total Sgr-related gamma-ray flux from 0 to $10^{-7}\;\mathrm{ph}\,\mathrm{cm}^{-2}\,\mathrm{s}^{-1}$. Since we apply the baseline model and hyper-parameter settings of Run 6, we also cover the case of imperfect background modelling. Indeed, as listed in the previous section, the gas component follows the spatial morphology of FGMA while the Run 6 spatial gas template is different. While this setup calls for a certain degree of background re-modulation, mock data and \sky~model coincide for some components, most notably the (structured) FBs and Sgr. 
    For each injected flux value, we generate 15 Poisson realisations to assess the statistical uncertainty of the recovered flux. At the same time, we quantify the significance $\mathcal{Z}_{\mathrm{Sgr}}$ as a function of the injected flux. The chosen range of gamma-ray fluxes encompasses the value obtained in \crocker. We report the results of this study in Sec.~\ref{sec:results-mock-case1}.
    \item[\textit{Case 2}:] Our objective is to examine the prospects of detecting an Sgr MSP population formed via the AIC mechanism. As demonstrated in Fig.~\ref{fig:sag_sim}, an MSP AIC population characterised by the total flux found in \crocker~may host a few MSPs that are around the detection threshold of the LAT. The bright part of the population would leave specific imprints in the residuals' or modulation parameters' structure when fitting with a smooth Sgr template as we do in our baseline setup. Therefore, we analyse mock data generated for the MSP AIC scenario with Run 8 hyper-parameters to, first,  asses the significance of Sgr and, second, to examine how the modulation parameters of the background components may be affected by the varying luminosity between the individual MSPs. We evaluate and interpret our results for this test in Sec.~\ref{sec:results-mock-case2}.
\end{itemize}

\subsection{Results: Case 1}
\label{sec:results-mock-case1}
\begin{figure*}
\centering\includegraphics[width=0.49\linewidth]{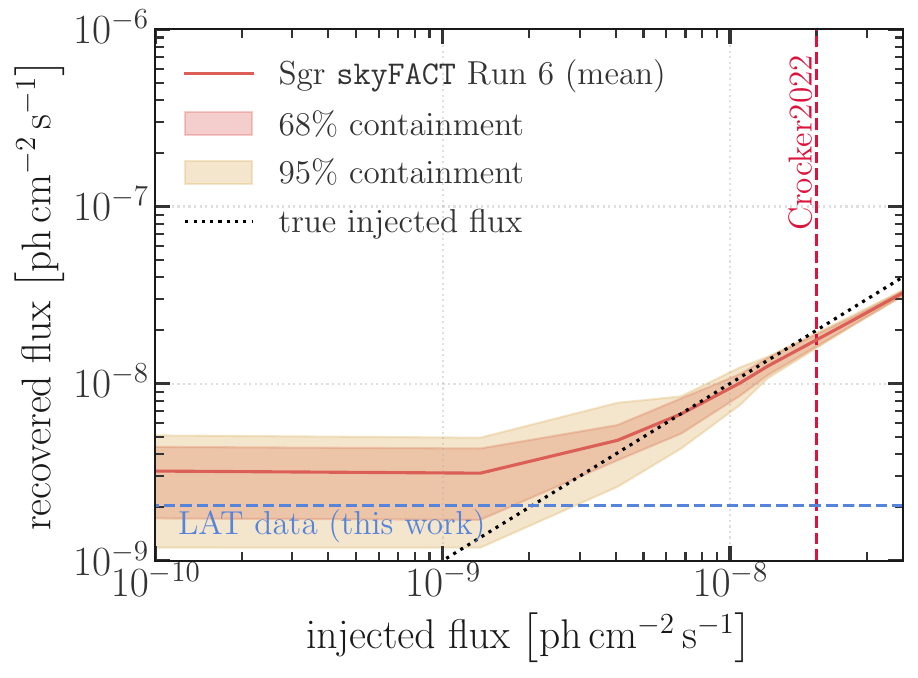}\hfill
    \includegraphics[width=0.48\linewidth]{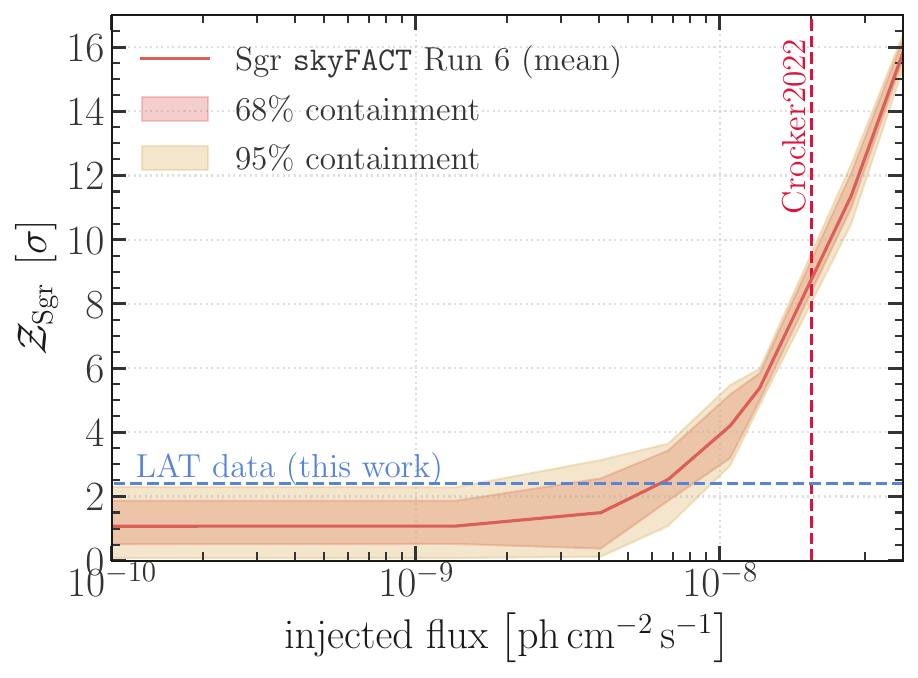}
    \caption{Results of the signal injection test conducted in the hyper-parameter settings of Run 6 and our baseline gamma-ray emission model regarding LAT mock data containing a purely diffuse MSP signal from Sgr. (\textit{Left}:) Evolution of the recovered flux of the Sgr component for a given flux of the injected Sgr diffuse emission (dotted black line). The solid red line marks the mean recovered flux derived from 15 realisations of the mock dataset while the red/orange bands denote the corresponding 68\%/95\% containment intervals. We show with a dashed blue line the obtained flux associated with the Sgr component when applying Run 6 to the real \textit{Fermi}-LAT data.  (\textit{Right}:) Significance $\mathcal{Z}_{\mathrm{Sgr}}$ of the Sgr component from comparison between a model without such a component. We adopt the colour code of the left panel. In both panels, we show as a vertical dashed red line the claimed flux value of Sgr derived in \protect\crocker. 
    \label{fig:sgr-diffuse-injection}}
\end{figure*}

A central question we can quantitatively answer is: Would \sky~significantly detect a diffuse MSP population in Sgr exhibiting a total flux of $2\times10^{-8}\;\mathrm{ph}\,\mathrm{cm}^{-2}\,\mathrm{s}^{-1}$ as reported in \crocker? 
Fig.~\ref{fig:sgr-diffuse-injection} displays the results we derived with the \sky~framework. The vertical, dashed, red line marks the claimed Sgr flux in \crocker. In the left panel, we compare the injected flux (dotted black line) to the recovered flux in terms of the mean from 15 Poisson realisations (solid red line) as well as its 68\% and 95\% containment bands. The putative value of $2\times10^{-8}\;\mathrm{ph}\,\mathrm{cm}^{-2}\,\mathrm{s}^{-1}$ falls into the region where the recovered flux follows the injected flux line. In fact, the significance $\mathcal{Z}_{\mathrm{Sgr}}$ of Sgr for this flux level is $\sim9\sigma$ as shown in the right panel of the same figure where we adopt the left panel's colour code. We note that the significance in \crocker~for the case of a structured FBs template (the same as here) is stated as $\sim8.1\sigma$, not far from the value we derived with our injection test. To summarise, if there were gamma-ray emissions following the baseline Sgr morphology at the magnitude that \crocker~report, we would have seen an unmistakably strong signal for the LAT dataset even in the \sky–approach. Yet, the horizontal, vertical, blue line in both panels shows the actual recovered flux and, thus, the significance of the baseline Run 6 findings on \textit{Fermi}-LAT data, which are well below this scenario. In our framework, the total Sgr emission is reduced by a factor of four or more with respect to \crocker. 

The injection test in the left panel of Fig.~\ref{fig:sgr-diffuse-injection} also demonstrates that the observed emission associated with Sgr in the LAT dataset is compatible with a pure upward fluctuation. The $1\sigma$ containment band for mock data with a faint Sgr component or even without Sgr emission encompasses the obtained Sgr total flux on real data (dashed blue line). However, the significance plot on the right points to a non-vanishing Sgr component with a flux $>6\times10^{-9}\;\mathrm{ph}\,\mathrm{cm}^{-2}\,\mathrm{s}^{-1}$.  
Despite the degree of mis-modelling we allowed for\footnote{\sky~model and mock data do not contain the same gas template.}, we stress that nonetheless, the injection test represents a rather well-controlled testing ground compared to the real data where mis-modelling may occur at an even higher level. This might also explain the minor inconsistency at the $2\sigma$ level between the injected flux consistent with the retrieved significance of Sgr in the LAT data and the recovered total flux. According to Fig.~\ref{fig:sgr-diffuse-injection}, such a significance should arise from an injected flux that is at least a factor of 1.5 larger than what we measured. However, this statement only applies to the $1\sigma$ level while both injection test and real data results are consistent at the $2\sigma$ level. We note that the displayed containment bands are not entirely converged due to the limited number of realisations, which result from the rather long computation time for each \sky~run.

Lastly, we point out that our injection test yields recovered flux values that systematically undershoot the injected fluxes when the associated Sgr component becomes quite significant, i.e.~fairly bright. While we only show results up to $4\times10^{-8}\;\mathrm{ph}\,\mathrm{cm}^{-2}\,\mathrm{s}^{-1}$, we checked that this effect persists up to $1\times10^{-7}\;\mathrm{ph}\,\mathrm{cm}^{-2}\,\mathrm{s}^{-1}$. Yet, the difference is never more pronounced than a factor of 1.5. In the case of a very bright Sgr region, the gas template absorbs part of its gamma-ray emission resulting in the reduced recovered flux level. Eventually, this observation does not limit the power of our argument since restoring the mis-attributed flux would most likely further increase the significance of an additional Sgr component and exacerbate the discrepancy between mock and real data.

\subsection{Results: Case 2}
\label{sec:results-mock-case2}


 As described in Section \ref{sec:sgr-msp-sim}, all population realisations of AIC MSPs reproduce the total gamma-ray flux for Sgr reported in \crocker. Therefore, our statements only refer to the plausibility of this scenario. In contrast to the previous section, we employ the baseline model and hyper-parameter settings of Run 8 to investigate the impact of the cumulative emission of point-like objects forming the injected signal on the fit residuals, or equivalently, the re-modulation parameters of the background components. We stress that the Sgr template in the \sky~model (used to fit the mock data) is still given by the (smooth) baseline spatial morphology without any freedom of spatial re-modulation. 

As a first remarkable result, we find that the spread of the significance of an additional Sgr component is dominated by the stochasticity of the MSPs' positions rather than the Poisson noise. The latter is the reason for the spread in the right panel of Fig.~\ref{fig:sgr-diffuse-injection}. Among ten examined Monte Carlo realisations of the population, we obtain the following extreme values for Sgr's significance: $\mathcal{Z}_{\mathrm{Sgr}, \mathrm{min}} = 0.35\sigma$ ($\sim29\%$ of injected flux recovered) and $\mathcal{Z}_{\mathrm{Sgr}, \mathrm{max}} = 5.71\sigma$ ($\sim92\%$ of injected flux recovered). Expressed differently, it depends very much on the exact MSP AIC population if it can be detected with \sky. From the set of realisations we tested, it appears that low and high levels of significance are equally likely. 

\begin{figure*}
\centering
\includegraphics[width=0.8\linewidth]{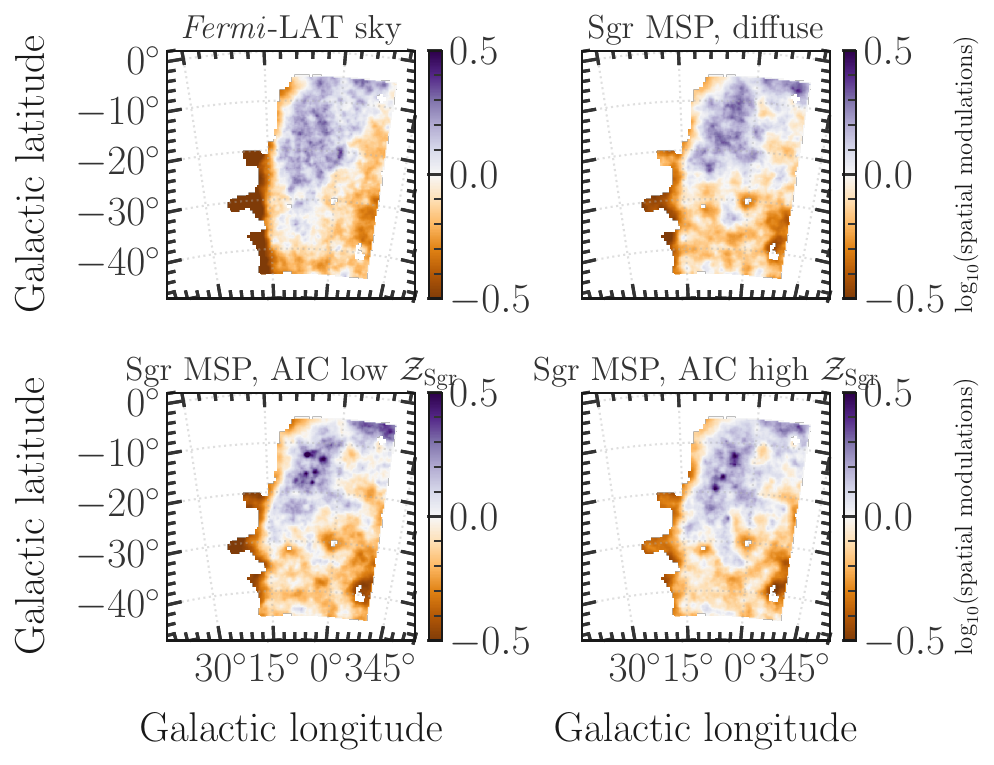}
    \caption{Map of the spatial modulation parameters $\bm{\tau}^{(\mathrm{FB})}$ for the FBs  (in log-scale) in the context of Run 8 hyper-parameters and our baseline gamma-ray emission model. We contrast four distinct cases: (\textit{upper left}:) \textit{Fermi}-LAT data, (\textit{upper right}:) mock data with a purely diffuse MSP population ({\it Case 1}) with a total flux corresponding to the value derived in \protect\crocker, (\textit{lower left}:) mock data prepared with an AIC MSP population ({\it Case 2}) resulting in the lowest obtained significance for Sgr and  (\textit{lower right}:) another AIC MSP realisation resulting in the highest obtained significance for Sgr.
    \label{fig:sgr-AIC-remodulation}}
\end{figure*}

What does this imply for our baseline Run 8 on real data? Going back to Tab.~\ref{tab:summary-baseline}, it states $\mathcal{Z}_{\mathrm{Sgr}, \mathrm{LAT}} = 0.7\sigma$ for LAT data. This seems compatible with the large spread of values we obtained for the AIC populations. In this light, there could still be a population hiding in the gamma-ray emission from Sgr. Yet, from the quoted values of recovered flux vs.~injected flux in the paragraph above, we would expect that the emission of such a putative population was mis-attributed to some of the \sky~model's background components. In Fig.~\ref{fig:sgr-AIC-remodulation}, we provide four maps of the spatial re-modulation parameters of the FBs template. The lower two panels show these parameters for the two extreme cases of the MSP AIC population with low and high significance mentioned above. The upper row's left panel contains the results for the \textit{Fermi}-LAT sky. In fact, we found that the FBs are the background contribution most degenerate with the Sgr component\footnote{We checked that the gas component absorbs a small but non-substantial fraction, too, without leaving a visible imprint on the re-modulation parameters.}. The FB modulation parameter maps are different just by eye. The AIC population realisations leave point-like features in the maps indicating that the FBs template absorbs residual photons exactly where bright MSPs, potentially around the detection threshold, are located. The difference between the low- and high-significance realisation is the number of bright MSPs. In contrast, the FBs' modulation parameters for the real data do not exhibit such pronounced point-like structures. It is a possibility, though, that we do not see similar features in the LAT panel because this part of Sgr's MSP population is already part of 4FGL-DR3. However, \crocker~used the same smooth Sgr template as us to derive the total gamma-ray flux associated with Sgr while also accounting for all known 4FGL sources at that time. In this sense, what we use as the total flux of Sgr is most likely not contaminated by emission from bright point-like sources. Seeing several bright MSPs in each realisation renders the AIC hypothesis less likely in itself. 

\begin{figure*}
\centering
\includegraphics[width=0.7\linewidth]{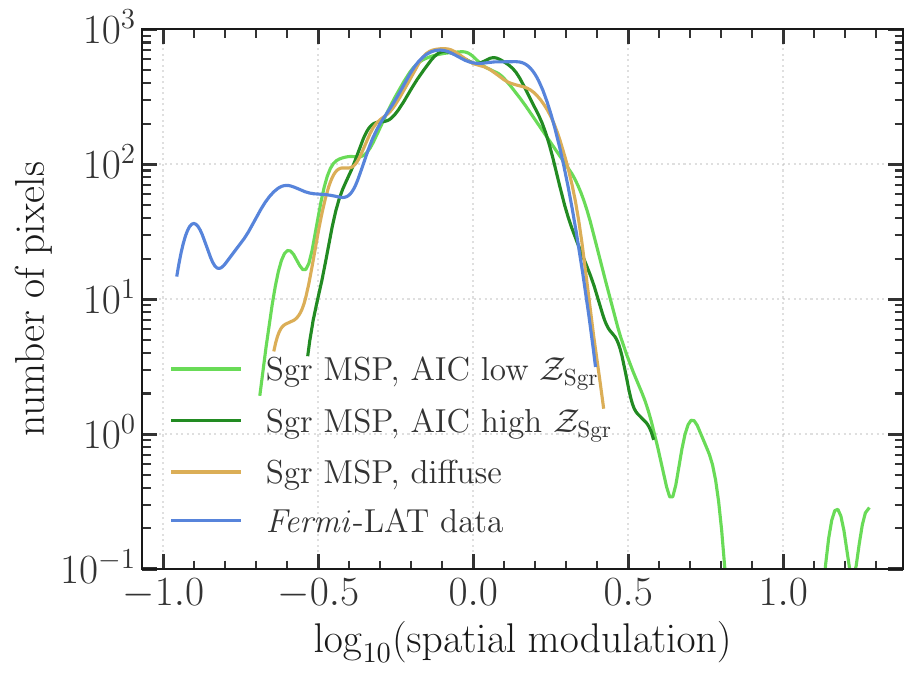}
    \caption{Histograms of the spatial Run-8 modulation parameters $\bm{\tau}^{(\mathrm{FB})}$ for the FBs shown in Fig.~\ref{fig:sgr-AIC-remodulation}. The modulation parameter profile associated with the real LAT data is shown in blue, the diffuse MSP population of {\it Case 1} is shown as an orange line while both AIC MSP realisations with high and low $\mathcal{Z}_{\mathrm{Sgr}}$ are displayed with two shades of green. Note that the provided histograms are slightly smoothed with a Gaussian kernel. 
    \label{fig:sgr-AIC-1pPDF-remodulation}}
\end{figure*}

In addition, we show in the right panel of the upper row in Fig.~\ref{fig:sgr-AIC-remodulation} how the FBs' modulation parameters are distributed for a purely diffuse MSP population ({\it Case 1}). There are no clear signs for bright point-like objects in this map similar to the results for the real LAT data. To make this comparison among the four tested scenarios more quantitative, we provide in Fig.~\ref{fig:sgr-AIC-1pPDF-remodulation} the one-point histograms of the spatial modulation parameters, i.e.~histograms that count how many pixels contain a certain value of the modulation parameters (for visual purposes these histograms are smoothed with a Gaussian kernel). \\
For small modulations around zero, all four scenarios perform comparably. The differences become evident for large downward and upward modulations. Regarding the downward modulations, the real LAT sky requires a much larger level than any other case. This is a direct consequence of the nature of our mock data. They contain the structured FBs template of \cite{Fermi-LAT:2014sfa}, which exhibits several ``arms'' towards positive longitude values (see Figs.~\ref{fig:sgr-AIC-remodulation} or \ref{fig:baseline-model}). Since also the flat FBs template allows for modulation of these arm structures, they are well reconstructed for mock data --  where they are present by definition -- but the real data do not seem to require them. \\
The more important difference is found for large upward modulations. The point-like features in the modulation maps are translated into longer tails of the histograms. In particular, the low-significance AIC population exhibits several pronounced features while the high-significance realisation follows up to a certain point. This is in contrast to the real LAT data, but also the purely diffuse population: They produce almost identical modulation parameter distributions with a lower cutoff than both extreme AIC cases. 

In summary, if there is gamma-ray emission from an MSP population in Sgr it must be almost 100\% diffuse. Yet, it cannot be at the level reported in \crocker~since, in that case, we should have detected it with a high significance (see Fig.~\ref{fig:sgr-diffuse-injection}). Alternatively, we may be in a somewhat contrived situation where the bright part of the Sgr AIC population is already contained in 4FGL-DR3 and there are essentially no MSPs close to the detection threshold such that the remaining population is purely diffuse.
However, this scenario is rather unlikely because:
(i) The luminosity function of the AIC population does not make sudden jumps and thereby evades a ``close-to-threshold'' part (see the histogram in Fig.~\ref{fig:sag_sim}).
 (ii) The total flux reported in \crocker~is mostly diffuse due to the limitations of template-based fits but at this flux level our Monte Carlo simulations always generate a few bright/detectable MSPs.
 (iii) If the rest of the population were purely diffuse, we would, again, expect a higher significance given the results of our injection test.
Therefore, we neither find evidence for an AIC MSP population in Sgr nor diffuse gamma-ray emission associated with Sgr at the level reported in \crocker.

\section{Conclusions}
\label{sec:conclusion}
In this study, we investigated the presence of gamma-ray emissions from the Sgr dSph in the so-called cocoon region of the FBs, as recently found by \crocker.
Adopting a hybrid approach which combines data-driven background optimisation methods (\sky) and photon-count statistical analyses (1pPDF), we put under stress
the evidence for Sgr emission, using 
various models and iterative runs. 
Our main findings are summarised as it follows:

\begin{itemize}
    \item We initially obtained strong evidence (\(\sim 14\sigma\)) for the presence of Sgr dSph in the cocoon region using a flat version of the FBs template, consistent with previous results by \crocker.
    \item The significance of Sgr dSph emission is highly sensitive to the modelling of the Galactic diffuse emission (including gas and IC maps) and the gamma-ray emission associated with the FBs. Optimising the FBs' structure from the initial flat template and allowing for spatial re-modulation of the foreground templates the significance of Sgr dSph emission is reduced to \(1.9\sigma\), and ultimately to approximately \(0.7\sigma\) in Run 8.
    \item A spectral analysis of the \sky~runs did not reveal a statistically significant component between 1 GeV and 10 GeV, the energy range where an MSP-like component would be most prominent. This supports the conclusion that the overall significance of the Sgr dSph emission is less than \(1\sigma\).
    \item We tested the impact of the initial model components' morphology on the significance of Sgr dSph emission using alternative templates. The results showed no significant evidence for Sgr dSph emission when accounting for systematics from the Sgr dSph stellar template, FBs, cocoon region, and diffuse Galactic emission components.
    \item An analysis using the 1pPDF revealed no significant differences in the \(dN/dS\) distribution between the Sgr and Anti-Sgr regions, indicating no additional source populations beyond the disc/extragalactic ones.
    \item Bayesian comparison of different gamma-ray diffuse emission models optimised with \sky~corroborated the findings, suggesting that the model from Run 8 provides a more accurate description of gamma rays within the Sgr dSph region than models showing significant evidence for diffuse Sgr dSph emission (e.g., Run 0).
    \item Through injection tests, we demonstrated that if gamma-ray emissions consistent with the baseline Sgr morphology and magnitude reported by \crocker~existed, we would have detected a strong signal in the \sky~approach. The observed emission is therefore consistent with a pure upward fluctuation.
    \item We also simulated the flux distribution of the putative MSPs population in the Sgr dSph of more than 600 MSPs following the AIC luminosity function, finding that they are expected to be very faint, and thus it will be very challenging to detect or disentangle them from the other components by inspecting the cumulative source-count distribution within the Sgr ROI. Thus, we find no evidence for an MSP population or diffuse gamma-ray emission from Sgr dSph at the level reported by \crocker.
\end{itemize}

In conclusion, after thorough iterations with \sky~and 1pPDF, we found no unequivocal evidence for gamma-ray emission from Sgr dSph or non-Poissonian variations indicative of a dim population of point-like sources. The data suggest that any observed emission is likely due to statistical fluctuations rather than a true astrophysical signal from the Sgr dSph.


\section*{Data Availability}

The \sky~and 1pPDF fit results, optimised model templates, generated MSP catalogues, etc.~produced in this analysis may be obtained upon reasonable request by contacting the authors. However, we want to stress that the re-modulation with \sky~produces optimised data-driven model components which have to be used with caution. For example, using re-modulated diffuse templates like the one for the gas or IC component is only meaningful within the model configuration employed to derive these templates in the first place. Making use of them in different contexts may naturally lead to mis-modelling and/or over-/under-fitting.

\begin{acknowledgments}
We thank F.~Donato and O.~Macias for their careful reading of the manuscript as well as their valuable comments. We also acknowledge useful discussions with D.~Malyshev and O.~Macias during the work on this analysis.
This work has been done thanks to the facilities offered by the Univ.~Savoie Mont Blanc - CNRS/IN2P3 MUST computing center.
CE acknowledges support by the ``Agence Nationale de la Recherche'', grant n.~ANR-19-CE31-0005-01 (PI: F.~Calore). The work of CE has been supported by the EOSC Future project which is co-funded by the European Union Horizon Programme call INFRAEOSC-03-2020, Grant Agreement 101017536. CE further acknowledges support from the COFUND action of Horizon Europe’s Marie Sk\l{}odowska-Curie Actions research programme, Grant Agreement 101081355 (SMASH).
SM acknowledges the European Union's Horizon Europe research and innovation program for support under the Marie Sklodowska-Curie Action HE MSCA PF–2021,  grant agreement No.10106280, project \textit{VerSi}. 
\end{acknowledgments}

\appendix

\section{Details of the baseline gamma-ray emission model for the Sgr dSph region}
\label{app:astro-model-details}

\begin{figure*}[t!]
\centering\includegraphics[width=0.99\linewidth]{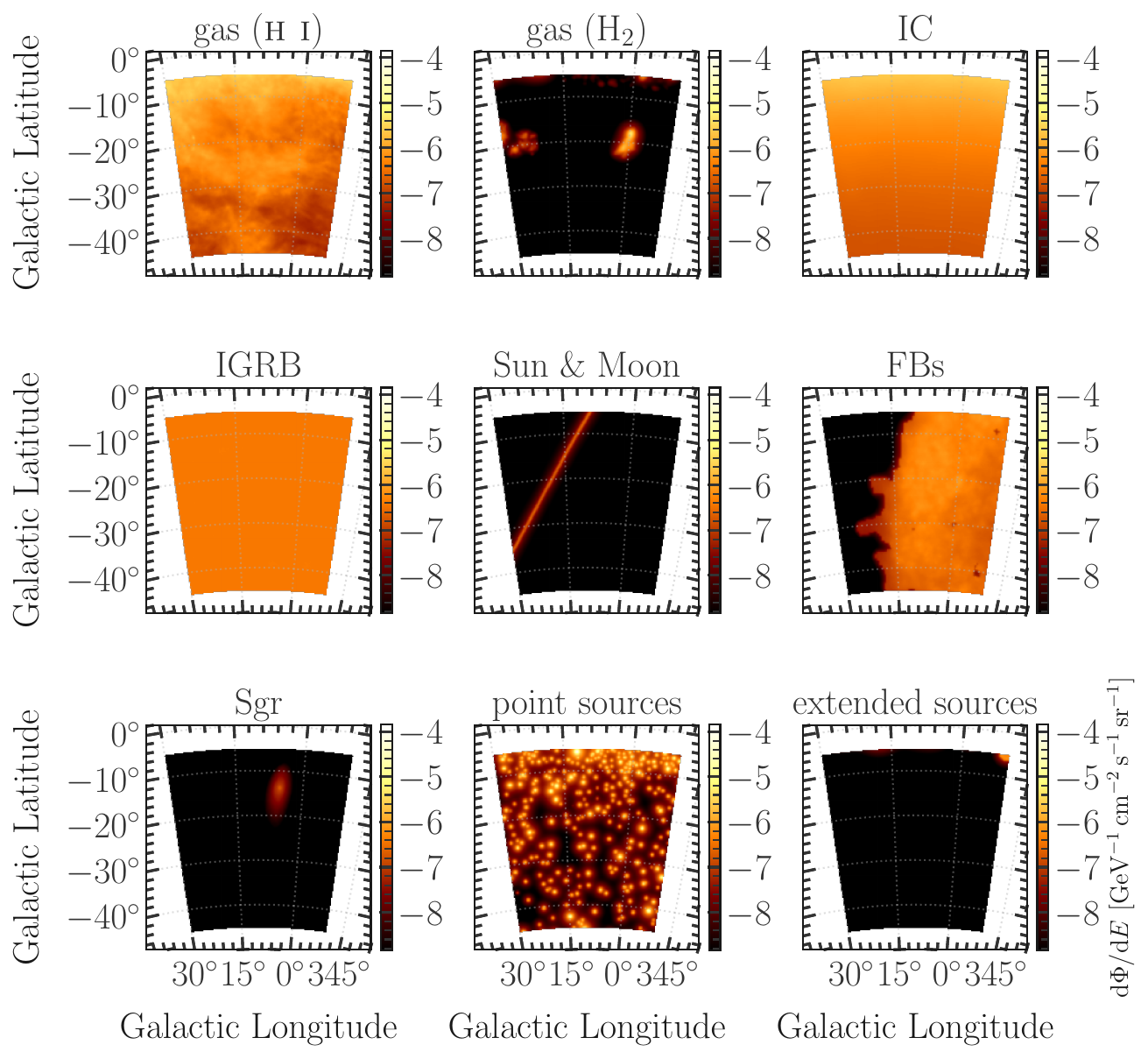}
    \caption{Summary of the spatial morphology of the baseline model's components shown in terms of their spectrum from 1 to 4 GeV after re-modulation with \sky~in Run 8. Note that all displayed components share the same colour bar. 
    \label{fig:baseline-model}}
\end{figure*}

All components of our baseline model setup are shown in Fig.~\ref{fig:baseline-model} in the energy band from 1 to 4 GeV as obtained after a \sky~fit with Run 8 hyper-parameters. In what follows, we provide more explanatory details about the model composition and the physics of the components.

\emph{(i) \& (ii)\;\;Point-like and extended sources.} As anticipated with the selection of the LAT dataset for the \sky~analysis part, we incorporate all 4FGL-DR3 sources\footnote{We utilised the catalogue version \href{https://fermi.gsfc.nasa.gov/ssc/data/access/lat/12yr_catalog/gll_psc_v30.fit}{\texttt{gll\_psc\_v30.fit}}.} residing within our ROI into the baseline gamma-ray emission model (629 sources in total). This concerns information about the sources' position (which is fixed to the nominal values report in 4FGL-DR3) and their spectra. An exception is those sources flagged as extended. We find the objects ``RX J1713.7-3946'', ``FHES J1741.6-3917'', ``FGES J1745.8-3028'', ``W28'', ``HESS J1804-216'', ``W30'', ``HESS J1808-204'', ``HESSJ 1809-193'', ``HESS J1825-137'', ``W41'', ``FGES J1838.9-0704'', ``HESS J1841-055'' and ``W44'' in our full ROI. All of them are located close to the Galactic equator and only mildly overlap with the designated fit region of our study. We incorporate them as individual components with spatial extension, i.e.~$T_p^{(k)}\neq0$, in our gamma-ray emission model. Each of them is initially modelled as a uniform brightness disc with a radial extension between $1.5^{\circ}$ to $3.0^{\circ}$ centred on their 4FGL-DR3 position.

\emph{(iii)\;\;IGRB.} The IGRB is incorporated as a fully isotropic component with a spectrum according to the chosen event class and type \texttt{iso\_P8R3\_ULTRACLEANVETO\_V3\_v1.txt}, which is part of the supplementary data of the \textit{Fermi} Science Tools. 

\emph{(iv)\;\;Sun\&Moon.} We adopt the two three-dimensional model cubes\footnote{This term usually refers to a flux model including the two-dimensional spatial morphology and spectrum in several energy bins.} for the emission from the Sun and Moon prepared by the authors of \crocker, which they made available in a zenodo repository \href{https://zenodo.org/records/6210967}{10.5281/zenodo.6210967}. Gamma-ray emission from the Moon is caused by Galactic cosmic rays triggering particle cascades in the direction of the Earth when hitting the lunar surface \cite{2012ApJ...758..140A}. A similar process occurs in the solar atmosphere while there is another possibility arising from IC scattering events of leptonic cosmic rays with the solar photon radiation field \cite{2011ApJ...734..116A}. The latter phenomenon generates an extended diffuse emission tracing the Sun's heliosphere. We note, however, that the adopted Sun and Moon templates are only correct in combination with the dataset from whose selection criteria (8-year 4FGL) they have been derived. Since we will eventually allow for spatial and spectral re-modulation of the emission, this initial input should provide a reasonable characterisation of the morphology.

\emph{(v)\;\;Gas.} The Galactic diffuse emission, produced via hadronic processes (mainly $\pi^0$ decay) and bremsstrahlung, originates in collisions of very-high-energy hadronic and leptonic cosmic rays with the interstellar medium/gas permeating the Milky Way. The eventually observed gamma-ray emission is proportional to the gas density and the injected flux of cosmic rays. While the created gamma rays may undergo further interactions with the interstellar medium and be subject to slight deflections from their point of origin (especially at energies beyond tens of TeV), the morphology of this component traces very well the distribution of gas. Therefore, we adopt in the baseline model setup two spatial templates that reflect the gas density of atomic hydrogen (\textsc{h~i}) taken from the 2016 release of the HI4PI collaboration \cite{2016A&A...594A.116H} and molecular hydrogen (H$_2$) represented by the third annulus (8.0 - 10.0 kpc) of the H$_2$ templates adopted in \crocker~and available on zenodo.org. We found that this annulus provides the sole sizeable contribution of H$_2$-induced gamma-ray emission. Moreover, we do not split our gas templates into multiple annuli as done in \crocker. It turns out that doing so increases the risk of losing the convexity property of the minimisation problem due to degeneracies among the annuli of the same gas class. The main focus of our analysis lies at Galactic latitudes far from the plane. Thus, we impose the spectrum of FGMA \cite{Fermi-LAT:2014ryh} as a prior on the spectrum of our baseline \textsc{h~i} and H$_2$ components. FGMA was devised to quantify the spectrum of the diffuse isotropic gamma-ray background in a work by the \fermi-LAT collaboration. We stress that it is not imperative to first process these gas density maps through a cosmic-ray propagation solver to obtain a prediction for the expected gamma-ray emission. The image reconstruction feature of \sky~should take care of such small- or large-scale deviations.

As alternative gas model (Model0) and cross-check, we adopt the set of gas templates utilised in \crocker, namely four annuli extend from 0--3.5~kpc, 3.5--8~kpc, 8--10~kpc and 10--50~kpc, in which the gas density has been determined through a hydrodynamical simulation of the Milky-Way gas content \cite{Pohl:2007dz}. On top of this ``luminous'' gas component, the authors also consider dust correction templates that account for ``dark gas'' or other inconsistencies in the reconstruction of the gas density from the raw measurements of molecular emission lines, as described in~\cite{Fermi-LAT:2019yla}. Besides, we also investigate the gas component of foreground model A (FGMA) \cite{Fermi-LAT:2014ryh}, which is based on different (older) observationally inferred gas densities.\footnote{The data files are publicly available at \url{https://www-glast.stanford.edu/pub_data/845/}. We extract FGMA's spectrum for our ROI accounting only for its $\pi^0$ decay and bremsstrahlung components.} We label it ModelA.
We impose the spectrum of FGMA \cite{Fermi-LAT:2014ryh} as a prior on the spectrum of our baseline \textsc{h~i} and H$_2$ components.

\emph{(vi)\;\;IC.} For the baseline, we represent the diffuse IC component by its spatial and spectral characterisation given by FGMA. This is one of the alternative IC models employed in \crocker. The IC emission is created by very-high-energy leptons scattering off of the low-energy ambient ISRFs -- predominantly the optical and infrared part -- of the Milky Way and the cosmic microwave background (CMB). The ISRF is to a large extent, and given the energies we are considering here, confined to regions of the Milky Way inhabited by stellar populations, i.e.~inside the Galactic disc. Since our analysis is conducted at higher latitudes, we do not expect the dominant part of the IC emission to stem from the ISRF-induced part but rather the one associated with the CMB. Above hundreds of GeV, i.e.~where our analysis is statistics-limited, the ISRF-induced component of the diffuse IC foreground may however yield the leading contribution. Therefore, we see a less pronounced need to split this component into multiple annuli as done in \crocker. 

As a variation of the baseline setup, we consider the original set of IC templates employed in \crocker, as alternative Model0. It is based on the IC component obtained in \cite{Johannesson:2018bit} using a three-dimensional model of the Milky-Way interstellar radiation fields (ISRFs) and the cosmic-ray propagation solver \texttt{GALPROP}. The IC emission is decomposed into the same four Galactocentric annuli as the gas emission.

\emph{(vii)\;\;FBs.} The FBs are an essential ingredient in our gamma-ray emission model since the cocoon region -- commonly attributed to this extended structure -- overlaps with the position of the Sgr dSph's core. Yet, the gamma-ray emission associated with the FBs does not seem to be related to a greater multi-wavelength picture of the region that would elucidate the origin of this structure. As a consequence, there is no reliable model of the FBs based on first principles (for attempts see, for example, \cite{Crocker:2014fla, Mertsch:2018ynu}) rendering it necessary to resort to data-driven derivations of their spatial and spectral profile. Such a data-driven model is essentially containing the residual emission of some sort of fit to the gamma-ray sky with known astrophysical emission components. As a decisive feature, the FBs exhibit a harder spectrum than any other astrophysical contribution at high latitudes rendering their identification possible. In return, this may bias our capabilities to assess the need for an additional gamma-ray emitter spatially coincident with the FBs as the additional component had already been absorbed into the FBs model provided that their spectral profile is degenerate enough. We adopt, as a baseline, the data-driven characterisation of the FBs in \cite{Fermi-LAT:2014sfa} in a twofold manner: For most of the iterative \sky~fit sequence, we model the FBs as a uniform source, i.e.~each pixel of the initial spatial template has the same value. We call this the \emph{flat} FBs template. Only in the last iterations, do we allow for structured emission from the FBs, i.e.~we take the data-driven model at face value. Both, flat and structured FBs, were probed in \crocker. In contrast to the spatial templates used for Sgr shown in Fig.~\ref{fig:sgr-templates} that display a single density peak towards Sgr's core, the structured FBs template exhibits many upward and downward fluctuations of the predicted gamma-ray emission. This is certainly the result of their data-driven derivation.
Additionally, we implement a more recent data-driven derivation of the low-latitude and high-latitude FBs from \cite{Fermi-LAT:2017opo}. In \cite{Macias:2019omb}, this map was improved with image reconstruction techniques. We adopt the latter spatial characterisation of the FBs as well as their spectrum derived in \cite{Fermi-LAT:2017opo} as alternative FB2017 model.\footnote{The files corresponding to the FBs template in Fig.~8 of Ref.~\cite{Fermi-LAT:2017opo} (two-component spectral analysis) are hosted at: \url{https://www-glast.stanford.edu/pub_data/1220/}}

\section{Full details on baseline runs}
\label{app:baseline-runs}

\subsection{Hyper-parameters definitions and results per run}
\label{app:baseline-full-table}

In this subsection, we provide the full details of each \sky~run conducted for the baseline setup, which was only partially shown in Tab.~\ref{tab:summary-baseline} in the main text. 
Besides the definition of the hyper-parameters for each model component, we also add the likelihood values obtained for each run with and without the Sgr template, as stated in Tab.~\ref{tab:summary-baseline-full}.

Going from Run 1 to Run 5, we start with the IGRB and only allow for about $5\%$ spectral variations. Afterwards and in the applied ordering, we allow for spatial and spectral re-modulation in the Sun+Moon, gas (\textsc{h i}), gas (H$_2$) and IC templates. We define similar hyper-parameters for the Sun+Moon and IC contribution motivated by the fact that the gamma-ray emission from the Sun is partially fueled by IC scattering events. This way, we ensure that the extended diffuse IC emission is treated consistently. The difference is a stronger constraint on the spatial modulation parameters of the Sun+Moon component. The spectrum may vary up to 25\%. Both gas templates are handled in the same manner. We leave generous freedom to the spatial modulation parameters but constrain the spectrum to remain within about 5\% of the model prediction to avoid the absorption of spatially overlapping gamma-ray components. In terms of smoothing scales, we neglect spectral smoothing but apply spatial smoothing to all diffuse components where the IC-related templates feature the largest spatial correlation length whereas the gas maps are only moderately smoothed. This distinction is motivated by the fact that local gas clumps can lead to small-scale enhancements of the gamma-ray emission, which is hard to model with the constraint of large correlation lengths.

\begin{table*}
    \centering
    {\renewcommand{\arraystretch}{1.5}
    \begin{tabular}{lcccccccccc}
    \Xhline{5\arrayrulewidth}
    \multirow{2}{*}{Components} & Run 0 & Run 1 & Run 2 & Run 3 & Run 4 & Run 5 & Run 6 & Run 7 & Run 8\\
     & \multicolumn{9}{c}{\sky~hyper-parameters:$\bigl[\begin{smallmatrix}
        \lambda & \lambda^{\prime} & \lambda^{\prime\prime} \\
        \eta & \eta^{\prime} & \cdot 
    \end{smallmatrix}\bigr]$}\\
    \hline
    4FGL-DR3 (PLS) & $\bigl[\begin{smallmatrix}
        \cdot & 25 & 10  \\
        \cdot & 0 & \cdot
    \end{smallmatrix}\bigr]$ & $\bigl[\begin{smallmatrix}
        \cdot & 25 & 10  \\
         \cdot & 0 & \cdot
    \end{smallmatrix}\bigr]$ & $\bigl[\begin{smallmatrix}
        \cdot & 25 & 10  \\
         \cdot & 0 & \cdot
    \end{smallmatrix}\bigr]$ & $\bigl[\begin{smallmatrix}
        \cdot & 25 & 10  \\
         \cdot & 0 & \cdot
    \end{smallmatrix}\bigr]$ & $\bigl[\begin{smallmatrix}
        \cdot & 25 & 10  \\
        \cdot & 0 & \cdot
    \end{smallmatrix}\bigr]$ & $\bigl[\begin{smallmatrix}
        \cdot & 25 & 10  \\
        \cdot & 0 & \cdot
    \end{smallmatrix}\bigr]$ & $\bigl[\begin{smallmatrix}
        \cdot & 25 & 10  \\
        \cdot & 0 & \cdot
    \end{smallmatrix}\bigr]$& $\bigl[\begin{smallmatrix}
        \cdot & 25 & 10  \\
        \cdot & 0 & \cdot
    \end{smallmatrix}\bigr]$ & $\bigl[\begin{smallmatrix}
        \cdot & 25 & 10  \\
        \cdot & 0 & \cdot
    \end{smallmatrix}\bigr]$ \\
    4FGL-DR3 (ext) & $\bigl[\begin{smallmatrix}
        0 & 1 & \infty  \\
        6 & 0 & \cdot
    \end{smallmatrix}\bigr]$ & $\bigl[\begin{smallmatrix}
        0 & 1 & \infty  \\
        6 & 0 & \cdot
    \end{smallmatrix}\bigr]$ & $\bigl[\begin{smallmatrix}
        0 & 1 & \infty  \\
         6 & 0 & \cdot
    \end{smallmatrix}\bigr]$ & $\bigl[\begin{smallmatrix}
        0 & 1 & \infty \\
         6 & 0 & \cdot
    \end{smallmatrix}\bigr]$ & $\bigl[\begin{smallmatrix}
        0 & 1 & \infty \\
         6 & 0 & \cdot
    \end{smallmatrix}\bigr]$ & $\bigl[\begin{smallmatrix}
        0 & 1 & \infty \\
        6 & 0 & \cdot
    \end{smallmatrix}\bigr]$ & $\bigl[\begin{smallmatrix}
        0 & 1 & \infty \\
        6 & 0 & \cdot
    \end{smallmatrix}\bigr]$ & $\bigl[\begin{smallmatrix}
        0 & 1 & \infty \\
        6 & 0 & \cdot
    \end{smallmatrix}\bigr]$ & $\bigl[\begin{smallmatrix}
        0 & 1 & \infty  \\
        6 & 0 & \cdot
    \end{smallmatrix}\bigr]$\\
    gas (\textsc{h i}) & $\bigl[\begin{smallmatrix}
        \infty & 0 & 0  \\
         0 & 0 & \cdot
    \end{smallmatrix}\bigr]$ & $\bigl[\begin{smallmatrix}
        \infty & 0 & 0 \\
         0 & 0 & \cdot
    \end{smallmatrix}\bigr]$ & $\bigl[\begin{smallmatrix}
        \infty &0& 0  \\
         0 & 0 & \cdot
    \end{smallmatrix}\bigr]$ & $\bigl[\begin{smallmatrix}
        \frac{1}{25} & 400 & 0  \\
         40 & 0 & \cdot
    \end{smallmatrix}\bigr]$ & $\bigl[\begin{smallmatrix}
        \frac{1}{25} & 400 & 0  \\
         40 & 0 & \cdot
    \end{smallmatrix}\bigr]$ & $\bigl[\begin{smallmatrix}
        \frac{1}{25} & 400 & 0  \\
         40 & 0 & \cdot
    \end{smallmatrix}\bigr]$ & $\bigl[\begin{smallmatrix}
        \frac{1}{25} & 400 & 0  \\
         40 & 0 & \cdot
    \end{smallmatrix}\bigr]$ & $\bigl[\begin{smallmatrix}
        \frac{1}{25} & 400 & 0  \\
         40 & 0 & \cdot
    \end{smallmatrix}\bigr]$ & $\bigl[\begin{smallmatrix}
        \frac{1}{25} & 44 & 0  \\
         40 & 0 & \cdot
    \end{smallmatrix}\bigr]$ \\
    gas (H$_2$) & $\bigl[\begin{smallmatrix}
        \infty & 0 & 0  \\
         0 & 0 & \cdot
    \end{smallmatrix}\bigr]$ & $\bigl[\begin{smallmatrix}
        \infty & 0 & 0 \\
         0 & 0 & \cdot
    \end{smallmatrix}\bigr]$ & $\bigl[\begin{smallmatrix}
        \infty &0& 0  \\
        0 & 0 & \cdot
    \end{smallmatrix}\bigr]$ & $\bigl[\begin{smallmatrix}
        \infty &0& 0  \\
         0 & 0 & \cdot
    \end{smallmatrix}\bigr]$ & $\bigl[\begin{smallmatrix}
        \frac{1}{25} & 400 & 0  \\
         40 & 0 & \cdot
    \end{smallmatrix}\bigr]$ & $\bigl[\begin{smallmatrix}
        \frac{1}{25} & 400 & 0  \\
         40 & 0 & \cdot
    \end{smallmatrix}\bigr]$ & $\bigl[\begin{smallmatrix}
        \frac{1}{25} & 400 & 0  \\
         40 & 0 & \cdot
    \end{smallmatrix}\bigr]$ & $\bigl[\begin{smallmatrix}
        \frac{1}{25} & 400 & 0  \\
        40 & 0 & \cdot
    \end{smallmatrix}\bigr]$ & $\bigl[\begin{smallmatrix}
        \frac{1}{25} & 44 & 0  \\
         40 & 0 & \cdot
    \end{smallmatrix}\bigr]$ \\
    IC & $\bigl[\begin{smallmatrix}
        \infty & 0 & 0  \\
        0 & 0 & \cdot
    \end{smallmatrix}\bigr]$ & $\bigl[\begin{smallmatrix}
        \infty & 0 & 0  \\
         0 & 0 & \cdot
    \end{smallmatrix}\bigr]$ & $\bigl[\begin{smallmatrix}
        \infty & 0 & 0  \\
         0 & 0 & \cdot
    \end{smallmatrix}\bigr]$ & $\bigl[\begin{smallmatrix}
        \infty & 0 & 0  \\
         0 & 0 & \cdot
    \end{smallmatrix}\bigr]$ & $\bigl[\begin{smallmatrix}
        \infty & 0 & 0  \\
         0 & 0 & \cdot
    \end{smallmatrix}\bigr]$ & $\bigl[\begin{smallmatrix}
        1 & 16 & 0  \\
         150 & 0 & \cdot
    \end{smallmatrix}\bigr]$ & $\bigl[\begin{smallmatrix}
        1 & 16 & 0  \\
        150 & 0 & \cdot
    \end{smallmatrix}\bigr]$ & $\bigl[\begin{smallmatrix}
        1 & 16 & 0  \\
        150 & 0 & \cdot
    \end{smallmatrix}\bigr]$ & $\bigl[\begin{smallmatrix}
        1 & 16 & 0  \\
        150 & 0 & \cdot
    \end{smallmatrix}\bigr]$\\
    IGRB & $\bigl[\begin{smallmatrix}
        \infty & 0 & 0  \\
         0 & 0 & \cdot
    \end{smallmatrix}\bigr]$ & $\bigl[\begin{smallmatrix}
        \infty & 400 & 0  \\
         0 & 0 & \cdot
    \end{smallmatrix}\bigr]$ & $\bigl[\begin{smallmatrix}
        \infty & 400 & 0  \\
         0 & 0 & \cdot
    \end{smallmatrix}\bigr]$ &  $\bigl[\begin{smallmatrix}
        \infty & 400 & 0  \\
         0 & 0 & \cdot
    \end{smallmatrix}\bigr]$ & $\bigl[\begin{smallmatrix}
        \infty & 400 & 0  \\
        0 & 0 & \cdot
    \end{smallmatrix}\bigr]$ & $\bigl[\begin{smallmatrix}
        \infty & 400 & 0  \\
        0 & 0 & \cdot
    \end{smallmatrix}\bigr]$ & $\bigl[\begin{smallmatrix}
        \infty & 400 & 0  \\
         0 & 0 & \cdot
    \end{smallmatrix}\bigr]$ & $\bigl[\begin{smallmatrix}
        \infty & 400 & 0  \\
        0 & 0 & \cdot
    \end{smallmatrix}\bigr]$ & $\bigl[\begin{smallmatrix}
        \infty & 400 & \frac{1}{25}  \\
        0 & 0 & \cdot
    \end{smallmatrix}\bigr]$\\
    Sun\&Moon & $\bigl[\begin{smallmatrix}
        \infty & 0 & 0  \\
         0 & 0 & \cdot
    \end{smallmatrix}\bigr]$ & $\bigl[\begin{smallmatrix}
        \infty & 0 & 0  \\
         0 & 0 & \cdot
    \end{smallmatrix}\bigr]$ & $\bigl[\begin{smallmatrix}
        10 & 16 & 0  \\
         150 & 0 & \cdot
    \end{smallmatrix}\bigr]$ & $\bigl[\begin{smallmatrix}
        10 & 16 & 0  \\
        150 & 0 & \cdot
    \end{smallmatrix}\bigr]$ & $\bigl[\begin{smallmatrix}
        10 & 16 & 0  \\
         150 & 0 & \cdot
    \end{smallmatrix}\bigr]$ & $\bigl[\begin{smallmatrix}
        10 & 16 & 0  \\
         150 & 0 & \cdot
    \end{smallmatrix}\bigr]$ & $\bigl[\begin{smallmatrix}
        10 & 16 & 0  \\
         150 & 0 & \cdot
    \end{smallmatrix}\bigr]$ & $\bigl[\begin{smallmatrix}
        10 & 16 & 0  \\
        150 & 0 & \cdot
    \end{smallmatrix}\bigr]$ & $\bigl[\begin{smallmatrix}
        10 & 16 & 0  \\
         150 & 0 & \cdot
    \end{smallmatrix}\bigr]$\\
    FBs (flat) & $\bigl[\begin{smallmatrix}
        \infty & 0 & 0  \\
         0 & 0 & \cdot
    \end{smallmatrix}\bigr]$ & $\bigl[\begin{smallmatrix}
        \infty & 0 & 0  \\
         0 & 0 & \cdot
    \end{smallmatrix}\bigr]$ & $\bigl[\begin{smallmatrix}
        \infty & 0 & 0  \\
         0 & 0 & \cdot
    \end{smallmatrix}\bigr]$ & $\bigl[\begin{smallmatrix}
        \infty & 0 & 0  \\
         0 & 0 & \cdot
    \end{smallmatrix}\bigr]$ & $\bigl[\begin{smallmatrix}
        \infty & 0 & 0  \\
         0 & 0 & \cdot
    \end{smallmatrix}\bigr]$ & $\bigl[\begin{smallmatrix}
        \infty & 0 & 0  \\
         0 & 0 & \cdot
    \end{smallmatrix}\bigr]$ & $-$ & $\bigl[\begin{smallmatrix}
        0 & 10^4 & \infty  \\
        6 & 0 & \cdot
    \end{smallmatrix}\bigr]$ & $\bigl[\begin{smallmatrix}
        0 & 400 & \frac{1}{25}  \\
         6 & 0 & \cdot
    \end{smallmatrix}\bigr]$ \\
    FBs (structured) & $-$ & $-$ & $-$ & $-$ & $-$ & $-$ & $\bigl[\begin{smallmatrix}
        \infty & 0 & 0  \\
         0 & 0 & \cdot
    \end{smallmatrix}\bigr]$ & $-$ & $-$\\
    Sgr & $\bigl[\begin{smallmatrix}
        \infty & 0 & 0  \\
         0 & 0 & \cdot
    \end{smallmatrix}\bigr]$ & $\bigl[\begin{smallmatrix}
        \infty & 0 & 0  \\
        0 & 0 & \cdot
    \end{smallmatrix}\bigr]$ & $\bigl[\begin{smallmatrix}
        \infty & 0 & 0  \\
         0 & 0 & \cdot
    \end{smallmatrix}\bigr]$ & $\bigl[\begin{smallmatrix}
        \infty & 0 & 0  \\
         0 & 0 & \cdot
    \end{smallmatrix}\bigr]$ & $\bigl[\begin{smallmatrix}
        \infty & 0 & 0  \\
         0 & 0 & \cdot
    \end{smallmatrix}\bigr]$ & $\bigl[\begin{smallmatrix}
        \infty & 0 & 0  \\
         0 & 0 & \cdot
    \end{smallmatrix}\bigr]$ & $\bigl[\begin{smallmatrix}
        \infty & 0 & 0  \\
         0 & 0 & \cdot
    \end{smallmatrix}\bigr]$ & $\bigl[\begin{smallmatrix}
        \infty & 0 & 0  \\
        0 & 0 & \cdot
    \end{smallmatrix}\bigr]$ & $\bigl[\begin{smallmatrix}
        \infty & 0 & 0  \\
        0 & 0 & \cdot
    \end{smallmatrix}\bigr]$\\
    \hline
    $-2\ln{\mathcal{L}}_{\mathrm{base}}$ & $309106$ & $309227$ & $309210$ & $297941$ & $297852 $ & $297648$ & $297439$ & $296224$ & $296146$ \\
    $-2\ln{\mathcal{L}}_{\mathrm{base+Sgr}}$ & $308879$ & $309013$ & $309002$ & $297881$ & $297780$ & $297584$ & $297417$ & $296207$ & $296137$\\
    $\mathcal{Z}_{\mathrm{Sgr}}\;\left[\sigma\right]$ & $13.6$ & $13.1$ & $12.9$ & $5.8$ & $6.5$ & $6.1$ & $2.4$ & $1.9$ & $0.7$\\
    \Xhline{5\arrayrulewidth}
    \end{tabular}}
    \caption{Summary of the iterative \sky-runs based on the baseline setup outlined in Sec.~\ref{sec:sgr_physics} completing the selected results presented in Tab.~\ref{tab:summary-baseline}. The notation is the same as in the latter. In addition, we provide the obtained log-likelihood values stating the sum of Poisson and penalising likelihood contribution for a fit with the baseline model ``\emph{base}'' and a model iteration also containing the Sgr dSph template- ``\emph{base+Sgr}''.
    \label{tab:summary-baseline-full}}
\end{table*}

\subsection{Fit residuals per run}
\label{app:residuals-per-run}

\begin{figure*}[t!]
\centering\includegraphics[width=0.99\linewidth]{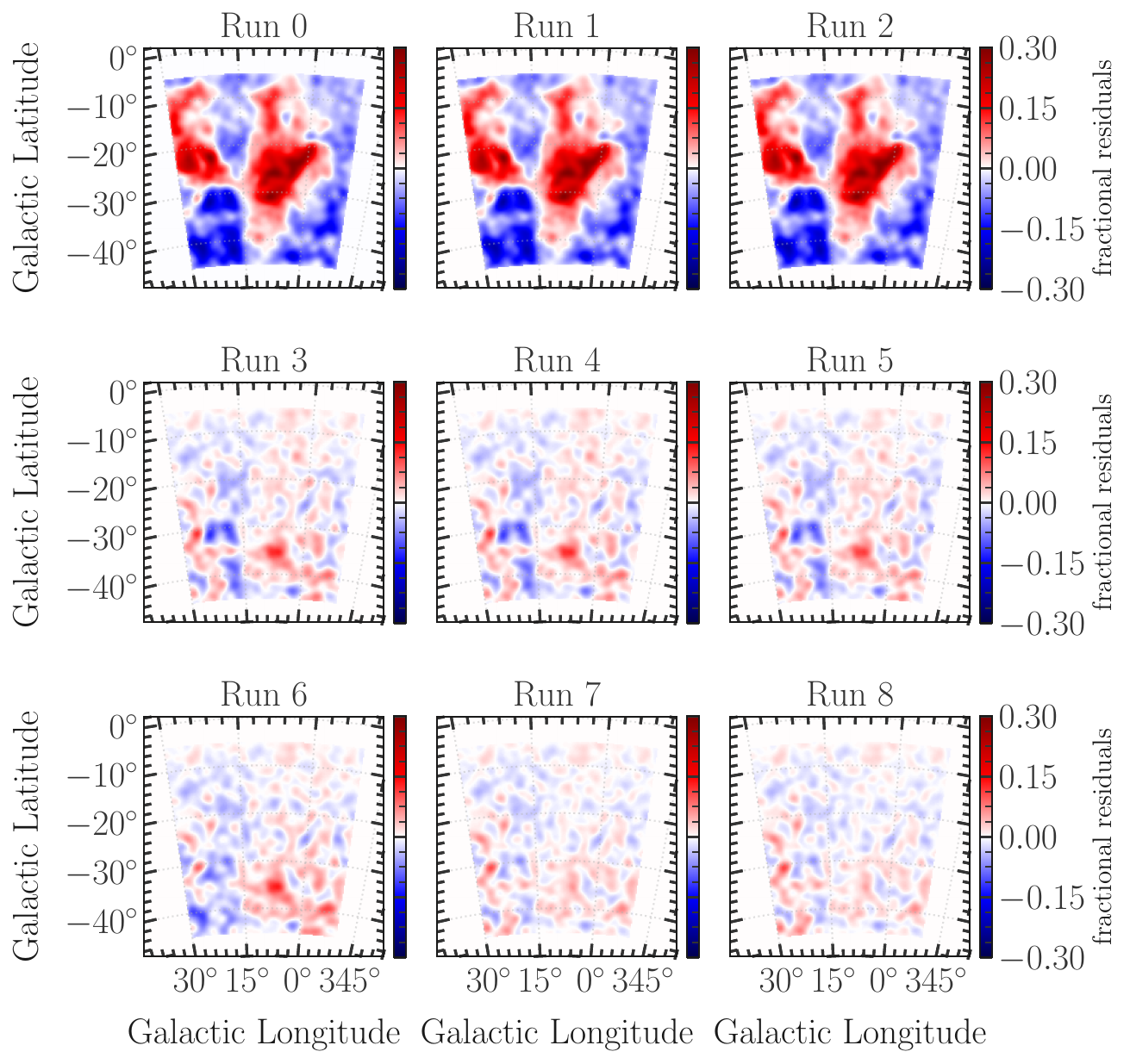}
    \caption{Same as the right panels in Fig.~\ref{fig:baseline-results} for each of the nine individual runs conducted for the baseline model setup. 
    \label{fig:baseline-all-residuals}}
\end{figure*}

To complement the baseline model's residuals for Run 0 and Run 8 shown in Fig.~\ref{fig:baseline-results} in the main text, we display the fit residuals for all runs of the baseline model setup in Fig.~\ref{fig:baseline-all-residuals}. All residual plots reflect the model setup with a template for the Sgr dSph. It becomes clear that the biggest improvement of the fit quality is the spatial modulation of the gas template(s) as demonstrated by the step from Run 2 to Run 3.

\subsection{Component re-modulation parameters per run}
\label{app:modulation-per-run}

\begin{figure*}[t!]
\centering\includegraphics[width=0.8\linewidth]{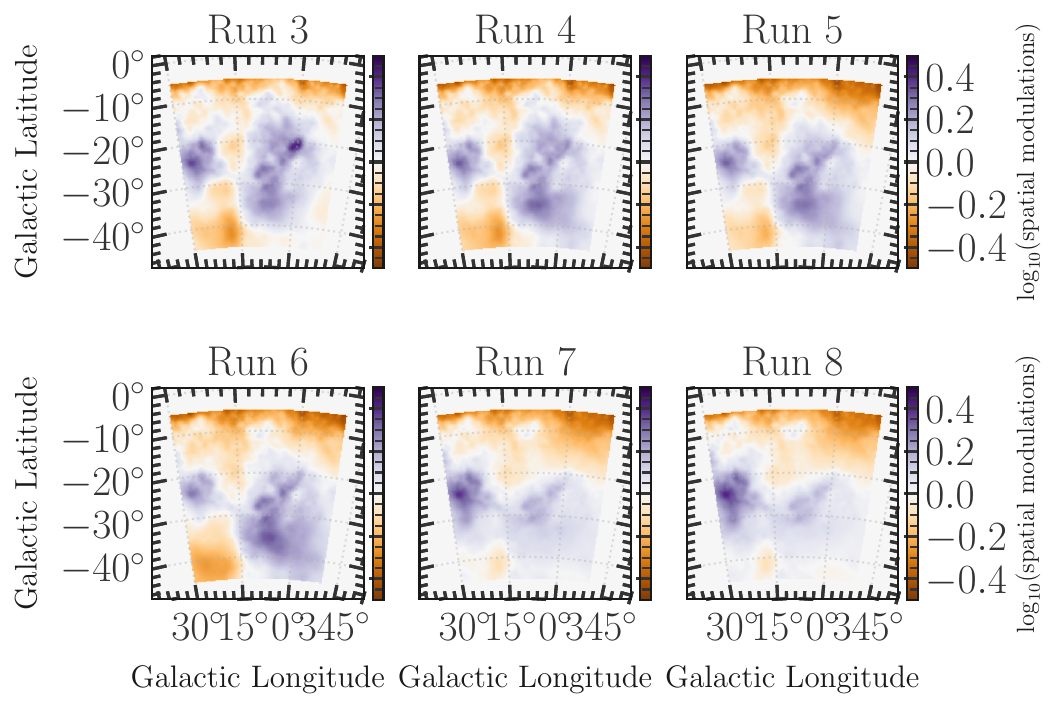}
    \caption{Same as the upper left panel of Fig.~\ref{fig:sgr-AIC-remodulation} regarding the \textsc{h i} gas template of the \sky~model's baseline setup including a Sgr component for all relevant runs discussed in Sec.~\ref{sec:results-skyfact} of the main text.  
    \label{fig:baseline-modulation-hi}}
\end{figure*}

\begin{figure*}[t!]
\centering\includegraphics[width=0.8\linewidth]{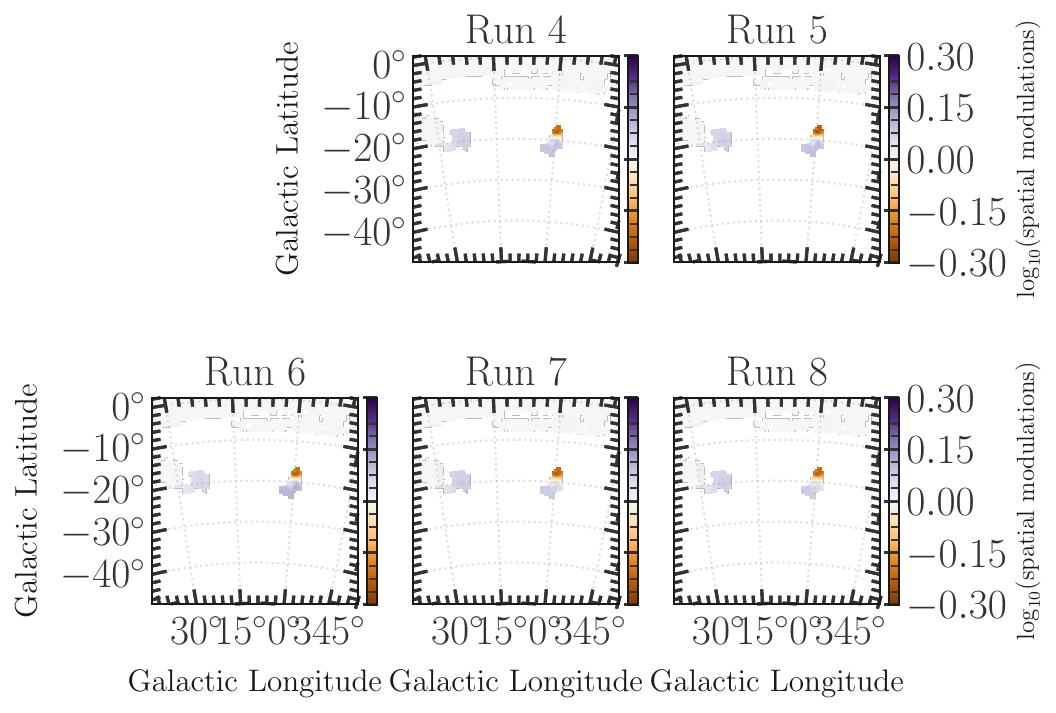}
    \caption{Same as Fig.~\ref{fig:baseline-modulation-hi} regarding the H$_2$ gas template of the \sky~model's baseline setup.  
    \label{fig:baseline-modulation-h2}}
\end{figure*}

\begin{figure*}[t!]
\centering\includegraphics[width=0.8\linewidth]{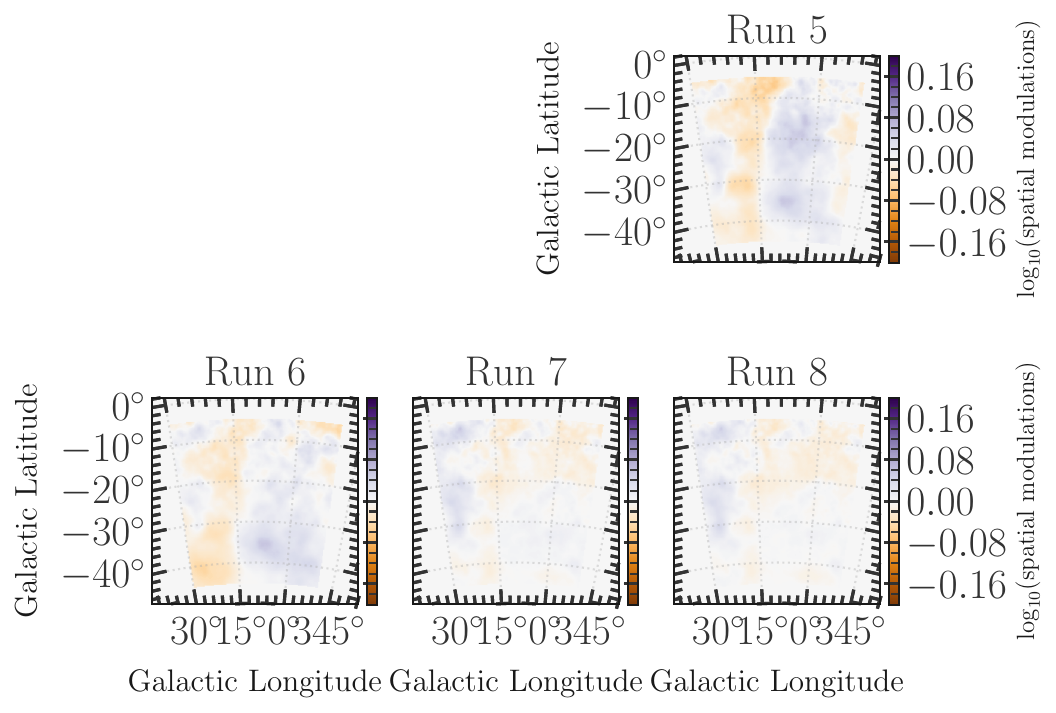}
    \caption{Same as Fig.~\ref{fig:baseline-modulation-hi} regarding the IC template of the \sky~model's baseline setup.  
    \label{fig:baseline-modulation-ic}}
\end{figure*}

\begin{figure*}[t!]
\centering\includegraphics[width=0.6\linewidth]{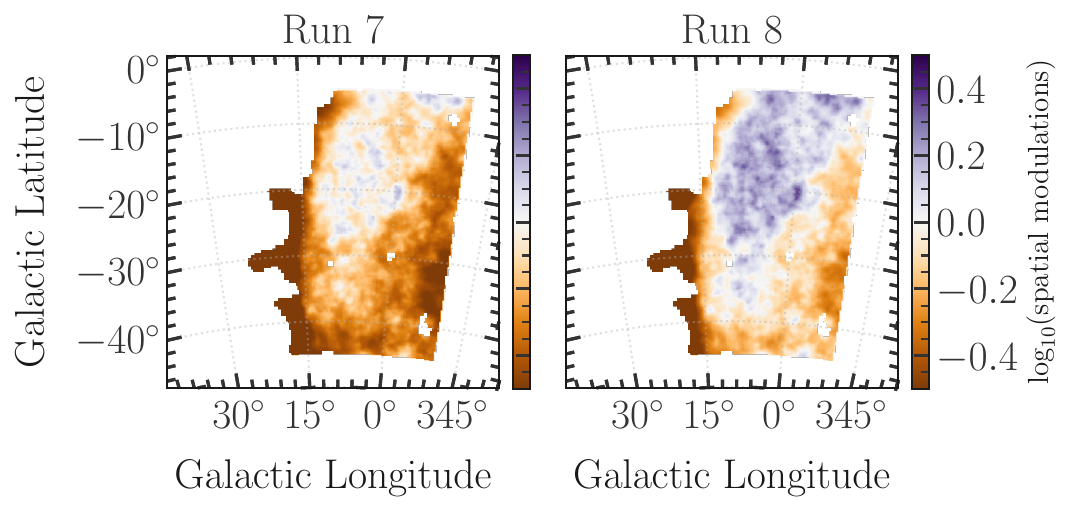}
    \caption{Same as Fig.~\ref{fig:baseline-modulation-hi} regarding the FBs template of the \sky~model's baseline setup.  
    \label{fig:baseline-modulation-fb}}
\end{figure*}

To visually show how \sky~modulates the spatial input templates, we provide in Figs.~\ref{fig:baseline-modulation-hi}, \ref{fig:baseline-modulation-h2}, \ref{fig:baseline-modulation-ic} and \ref{fig:baseline-modulation-fb} the decadic logarithm of the spatial re-modulation parameters of the gas (\textsc{h i}), gas (H$_2$), IC and FBs components for all relevant runs of the baseline model setup in analogy to the plot in Fig.~\ref{fig:sgr-AIC-remodulation}. These model components have the largest impact on the significance of Sgr, which was part of the fit model.

\section{Full details for the alternative runs of ModelA}
\label{app:modelA-runs}

In this section, we present in Tab.~\ref{tab:summary-alternative-modelA} the full details of hyper-parameter definition and results for the alternative iterative \sky~runs using ModelA as the basis for the gas component which we discussed in Sec.~\ref{sec:baseline-robustness} of the main text. The summary table follows the notation established for Tab.~\ref{tab:summary-baseline-full}.

\begin{table*}
    \centering
    {\renewcommand{\arraystretch}{1.5}
    \begin{tabular}{lccccccc}
    \Xhline{5\arrayrulewidth}
    \multirow{2}{*}{Components} & Run 2 & Run 3 + 4 & Run 5 & Run 6 & Run 7 & Run 8\\
     & \multicolumn{6}{c}{\sky~hyper-parameters:$\bigl[\begin{smallmatrix}
        \lambda & \lambda^{\prime} & \lambda^{\prime\prime} \\
        \eta & \eta^{\prime} & \cdot 
    \end{smallmatrix}\bigr]$}\\
    \hline
    4FGL-DR3 (PLS) & $\bigl[\begin{smallmatrix}
        \cdot & 25 & 10  \\
         \cdot & 0 & \cdot
    \end{smallmatrix}\bigr]$ & $\bigl[\begin{smallmatrix}
        \cdot & 25 & 10  \\
        \cdot & 0 & \cdot
    \end{smallmatrix}\bigr]$ & $\bigl[\begin{smallmatrix}
        \cdot & 25 & 10  \\
        \cdot & 0 & \cdot
    \end{smallmatrix}\bigr]$ & $\bigl[\begin{smallmatrix}
        \cdot & 25 & 10  \\
        \cdot & 0 & \cdot
    \end{smallmatrix}\bigr]$& $\bigl[\begin{smallmatrix}
        \cdot & 25 & 10  \\
        \cdot & 0 & \cdot
    \end{smallmatrix}\bigr]$ & $\bigl[\begin{smallmatrix}
        \cdot & 25 & 10  \\
        \cdot & 0 & \cdot
    \end{smallmatrix}\bigr]$ \\
    4FGL-DR3 (ext) & $\bigl[\begin{smallmatrix}
        0 & 1 & \infty  \\
         6 & 0 & \cdot
    \end{smallmatrix}\bigr]$ & $\bigl[\begin{smallmatrix}
        0 & 1 & \infty \\
         6 & 0 & \cdot
    \end{smallmatrix}\bigr]$ & $\bigl[\begin{smallmatrix}
        0 & 1 & \infty \\
        6 & 0 & \cdot
    \end{smallmatrix}\bigr]$ & $\bigl[\begin{smallmatrix}
        0 & 1 & \infty \\
        6 & 0 & \cdot
    \end{smallmatrix}\bigr]$ & $\bigl[\begin{smallmatrix}
        0 & 1 & \infty \\
        6 & 0 & \cdot
    \end{smallmatrix}\bigr]$ & $\bigl[\begin{smallmatrix}
        0 & 1 & \infty  \\
        6 & 0 & \cdot
    \end{smallmatrix}\bigr]$\\
    gas (\textsc{h i} + H$_2$, FGMA) & $\bigl[\begin{smallmatrix}
        \infty &0& 0  \\
         0 & 0 & \cdot
    \end{smallmatrix}\bigr]$ & $\bigl[\begin{smallmatrix}
        \frac{1}{25} & 400 & 0  \\
         40 & 0 & \cdot
    \end{smallmatrix}\bigr]$ & $\bigl[\begin{smallmatrix}
        \frac{1}{25} & 400 & 0  \\
         40 & 0 & \cdot
    \end{smallmatrix}\bigr]$ & $\bigl[\begin{smallmatrix}
        \frac{1}{25} & 400 & 0  \\
         40 & 0 & \cdot
    \end{smallmatrix}\bigr]$ & $\bigl[\begin{smallmatrix}
        \frac{1}{25} & 400 & 0  \\
         40 & 0 & \cdot
    \end{smallmatrix}\bigr]$ & $\bigl[\begin{smallmatrix}
        \frac{1}{25} & 44 & 0  \\
         40 & 0 & \cdot
    \end{smallmatrix}\bigr]$ \\
    IC & $\bigl[\begin{smallmatrix}
        \infty & 0 & 0  \\
         0 & 0 & \cdot
    \end{smallmatrix}\bigr]$ & $\bigl[\begin{smallmatrix}
        \infty & 0 & 0  \\
         0 & 0 & \cdot
    \end{smallmatrix}\bigr]$ & $\bigl[\begin{smallmatrix}
        1 & 16 & 0  \\
         150 & 0 & \cdot
    \end{smallmatrix}\bigr]$ & $\bigl[\begin{smallmatrix}
        1 & 16 & 0  \\
        150 & 0 & \cdot
    \end{smallmatrix}\bigr]$ & $\bigl[\begin{smallmatrix}
        1 & 16 & 0  \\
        150 & 0 & \cdot
    \end{smallmatrix}\bigr]$ & $\bigl[\begin{smallmatrix}
        1 & 16 & 0  \\
        150 & 0 & \cdot
    \end{smallmatrix}\bigr]$\\
    IGRB & $\bigl[\begin{smallmatrix}
        \infty & 400 & 0  \\
         0 & 0 & \cdot
    \end{smallmatrix}\bigr]$ & $\bigl[\begin{smallmatrix}
        \infty & 400 & 0  \\
        0 & 0 & \cdot
    \end{smallmatrix}\bigr]$ & $\bigl[\begin{smallmatrix}
        \infty & 400 & 0  \\
        0 & 0 & \cdot
    \end{smallmatrix}\bigr]$ & $\bigl[\begin{smallmatrix}
        \infty & 400 & 0  \\
         0 & 0 & \cdot
    \end{smallmatrix}\bigr]$ & $\bigl[\begin{smallmatrix}
        \infty & 400 & 0  \\
        0 & 0 & \cdot
    \end{smallmatrix}\bigr]$ & $\bigl[\begin{smallmatrix}
        \infty & 400 & \frac{1}{25}  \\
        0 & 0 & \cdot
    \end{smallmatrix}\bigr]$\\
    Sun\&Moon & $\bigl[\begin{smallmatrix}
        10 & 16 & 0  \\
         150 & 0 & \cdot
    \end{smallmatrix}\bigr]$ & $\bigl[\begin{smallmatrix}
        10 & 16 & 0  \\
         150 & 0 & \cdot
    \end{smallmatrix}\bigr]$ & $\bigl[\begin{smallmatrix}
        10 & 16 & 0  \\
         150 & 0 & \cdot
    \end{smallmatrix}\bigr]$ & $\bigl[\begin{smallmatrix}
        10 & 16 & 0  \\
         150 & 0 & \cdot
    \end{smallmatrix}\bigr]$ & $\bigl[\begin{smallmatrix}
        10 & 16 & 0  \\
        150 & 0 & \cdot
    \end{smallmatrix}\bigr]$ & $\bigl[\begin{smallmatrix}
        10 & 16 & 0  \\
         150 & 0 & \cdot
    \end{smallmatrix}\bigr]$\\
    FBs (flat) & $\bigl[\begin{smallmatrix}
        \infty & 0 & 0  \\
         0 & 0 & \cdot
    \end{smallmatrix}\bigr]$ & $\bigl[\begin{smallmatrix}
        \infty & 0 & 0  \\
         0 & 0 & \cdot
    \end{smallmatrix}\bigr]$ & $\bigl[\begin{smallmatrix}
        \infty & 0 & 0  \\
         0 & 0 & \cdot
    \end{smallmatrix}\bigr]$ & $-$ & $\bigl[\begin{smallmatrix}
        0 & 10^4 & \infty  \\
        6 & 0 & \cdot
    \end{smallmatrix}\bigr]$ & $\bigl[\begin{smallmatrix}
        0 & 400 & \frac{1}{25}  \\
         6 & 0 & \cdot
    \end{smallmatrix}\bigr]$ \\
    FBs (structured) & $-$ & $-$ & $-$ & $\bigl[\begin{smallmatrix}
        \infty & 0 & 0  \\
         0 & 0 & \cdot
    \end{smallmatrix}\bigr]$ & $-$ & $-$\\
    Sgr & $\bigl[\begin{smallmatrix}
        \infty & 0 & 0  \\
         0 & 0 & \cdot
    \end{smallmatrix}\bigr]$ & $\bigl[\begin{smallmatrix}
        \infty & 0 & 0  \\
         0 & 0 & \cdot
    \end{smallmatrix}\bigr]$ & $\bigl[\begin{smallmatrix}
        \infty & 0 & 0  \\
         0 & 0 & \cdot
    \end{smallmatrix}\bigr]$ & $\bigl[\begin{smallmatrix}
        \infty & 0 & 0  \\
         0 & 0 & \cdot
    \end{smallmatrix}\bigr]$ & $\bigl[\begin{smallmatrix}
        \infty & 0 & 0  \\
        0 & 0 & \cdot
    \end{smallmatrix}\bigr]$ & $\bigl[\begin{smallmatrix}
        \infty & 0 & 0  \\
        0 & 0 & \cdot
    \end{smallmatrix}\bigr]$\\
    \hline
    $-2\ln{\mathcal{L}}_{\mathrm{base}}$ & $303277$ & $296032$ & $296097$ & $296035$ & $295090$ & $295069$ \\
    $-2\ln{\mathcal{L}}_{\mathrm{base+Sgr}}$ & $303121$ & $295961$ & $296038$ & $296009$ & $295074$ & $295055$\\
    $\mathcal{Z}_{\mathrm{Sgr}}\;\left[\sigma\right]$ & $10.8$ & $6.5$ & $5.7$ & $2.9$ & $1.7$ & $1.5$\\
    \Xhline{5\arrayrulewidth}
    \end{tabular}}
    \caption{Same as Tab.~\ref{tab:summary-baseline-full} summarising the results obtained in the case of the alternative diffuse initial templates of ModelA. 
    \label{tab:summary-alternative-modelA}}
\end{table*}

\clearpage
\bibliographystyle{bib_style}
\bibliography{sgrdwarf_bib}

\end{document}